\documentclass[prd,twocolumn,superscriptaddress,amsmath,showpacs,amssymb,floatfix,nofootinbib,10pt]{revtex4}
\usepackage{indentfirst}
\usepackage{tikz}
\usepackage{setspace}
\usepackage{ulem,amsfonts,color,xcolor,graphicx,amsfonts,multirow,amsmath}
\usepackage[section]{placeins}
\definecolor{bluee}{rgb}{0,0,1}
\usepackage[dvipdfm, colorlinks=true, colorlinks=true, citecolor=bluee, linkcolor=bluee, urlcolor=bluee]{hyperref}

\allowdisplaybreaks

\newcommand{\hh}{\hspace{0.25cm}}
\newcommand{\colorr}[1]{{\color{blue}#1}}
\begin{document}

\title{Scrutinizing $B^0$-meson flavor changing neutral current decay into scalar $K_0^*(1430)$ meson with $b\to s \ell^+\ell^-(\nu\bar\nu)$ transition}
\author{Yin-Long Yang$^*$}
\author{Ya-Xiong Wang\footnote{Yin-Long Yang and Ya-Xiong Wang contributed equally to this work.}}
\affiliation{Department of Physics, Guizhou Minzu University, Guiyang 550025, P.R.China}
\author{Hai-Bing Fu}
\email{fuhb@gzmu.edu.cn}
\affiliation{Department of Physics, Guizhou Minzu University, Guiyang 550025, P.R.China}
\affiliation{Institute of High Energy Physics, Chinese Academy of Sciences, Beijing 100049, P.R.China}
\author{Tao Zhong}
\author{Ya-Lin Song}
\affiliation{Department of Physics, Guizhou Minzu University, Guiyang 550025, P.R.China}

\begin{abstract}
In this paper, we investigate the rare decay $B^0\to K_0^*(1430)\ell^+\ell^-$ with $\ell=(e,\mu,\tau)$ and $B^0\to K_0^*(1430)\nu\bar{\nu}$ induced by the flavor changing neutral current transition of $b\to s\ell^+\ell^-(\nu\bar\nu)$. Firstly, the $B^0\to K_0^*(1430)$ transition form factors (TFFs) are calculated by using the QCD light-cone sum rule approach up to next-to-leading order accuracy. In which the $K_0^*(1430)$-meson twist-2 and twist-3 LCDAs have been calculated both from the SVZ sum rule in the background field theory framework and light-cone harmonic oscillator model. Then, we obtained the three TFFs at large recoil point, {\it i.e.,} $f_+^{B^0\to K_0^\ast}(0)= 0.470_{-0.101}^{+0.086}$, $f_-^{B^0\to K_0^\ast}(0)= -0.340_{-0.068}^{+0.068}$, and $f_{\rm T}^{B^0\to K_0^\ast}(0)= 0.537^{+0.112}_{-0.115}$. Meanwhile, we extrapolated TFFs to the whole physical $q^2$-region by using the simplified $z(q^2)$-series expansion. Furthermore, we calculate the $B^0\to K_0^*(1430)\ell^+\ell^-(\nu\bar{\nu})$ decay widths, branching fractions, and longitudinal lepton polarization asymmetries of $B^0\to K_0^*(1430)\ell^+\ell^-$, which lead to ${\cal B}(B^0\to K_0^*(1430)e^+e^-) = (6.65^{+2.52}_{-2.42})\times 10^{-7}$, ${\cal B}(B^0\to K_0^*(1430)\mu^+\mu^-)=(6.62^{+2.51}_{-2.41})\times 10^{-7}$, ${\cal B}(B^0\to K_0^*(1430)\tau^+\tau^-)=(1.88^{+1.10}_{-0.97})\times 10^{-8}$, ${\cal B}(B^0\to K_0^*(1430)\nu\bar{\nu})= 3.85^{+1.55}_{-1.48}\times 10^{-6}$ and the integrated longitudinal lepton polarization asymmetries $\langle A_{P_L} \rangle = (-0.99, -0.96, -0.03)$ for the cases $\ell=(e, \mu, \tau)$ respectively.
\end{abstract}
\date{\today}
\maketitle
\newpage

\section{Introduction}
Based on the flavour changing neutral current (FCNC) processes, the beauty decays induced by $b\to s\ell^+\ell^-(\ell=e,\mu,\tau)$ transitions have gained much attention over the last decades, both from theoretical and experimental fronts. The FCNC decay processes are highly suppressed in the Standard Model (SM) at tree level, but they can be induced at the loop level by Glashow-Iliopoulos-Miani (GIM) amplitudes~\cite{Glashow:1970gm, Ahmed:2016jgv}. This renders it the most sensitive test for the SM, and its branching fractions are particularly sensitive to the presence of virtual particles, which are predicted to arise in extensions of the SM~\cite{LHCb:2014vgu}. Moreover, another reason why these decay processes are suppressed in SM is that they rely on the weak mixing angle of the quark-flavor rotation Cabibbo-Kobayashi-Maskawa (CKM) matrix elements~\cite{Kobayashi:1973fv}. The above cases make the FCNC processes relatively rare. In these decay processes, there exist many interesting observables that can be probed and discussed, such as angular observables, hadronic mass spectrum, leptonic forward-backward asymmetries, lepton polarization asymmetries, and lepton flavor universality (LFU) ratios~\cite{Hurth:2014vma, Das:2018sms,LHCb:2013ghj, Huber:2023qse, Sato:2013eva, Lunghi:2010tr, Mahata:2022cxf, Falahati:2015hwa}. These observables can provide a reliable scope for gaining a deeper understanding of how to handle the SM uncertainties. Thus, it is considered well suited to test the SM at one loop level in flavor physics and offer valuable information to determine the CKM matrix elements $|V_{tb}|$ and $|V_{ts}|$.

Experimentally, the rare $B$-meson decay was reported by the CLEO Collaboration several decades ago, which is discussed through exclusive decay $B\to K^\ast \gamma$ and inclusive decay $B\to X_s\gamma$, respectively~\cite{CLEO:1993nic,CLEO:1994veu}. These processes prompt a lot of theoretical interest on rare $B$-meson decay, and open a new window to explore new physics. Following this hot spot, the Belle Collaboration reported the first observation of the FCNC weak decay process $B\to K\ell^+\ell^-$ with $\ell=(e,\mu)$ in 2001~\cite{Belle:2001oey}. They used the data sample of 29.1 fb$^{-1}$ accumulated at the $\Upsilon(4S)$ resonance and contains 31.3 million $B\bar{B}$ pairs to yield the result of the branching fraction $\mathcal{B} (B\to K\ell^+\ell^-)$, which is expected to be $(0.75^{+0.25}_{-0.21}\pm 0.09)\times 10^{-6}$. Then, BaBar Collaboration presented evidence for FCNC processes $B\to K(K^\ast)\ell^+\ell^-$ with branching fraction $\mathcal{B}(B\to K \ell^+\ell^-)= (0.65^{+0.14}_{-0.13}\pm 0.04)\times 10^{-6}$, which is in agreement with the above Belle result~\cite{BaBar:2003szi}. The recent LHCb Collaboration~\cite{LHCb:2022vje} in 2022 studied $B^+(B^0)\to K^+(K^{*0})\ell^+\ell^-$ decays and presented the precise results from the first simultaneous test of LFU for these two decay processes in the low-$q^2$ and central-$q^2$-regions, which is also a powerful null test of the SM. Up to now, the $b\to s\ell^+\ell^-$ process has been explored by many experimental groups, such as CDF~\cite{CDF:2011grz,CDF:2011tds}, Belle~\cite{Belle:2008knz,Belle:2005hjc,Belle:2009zue,Belle:2021ecr,Belle:2019oag}, BaBar~\cite{BaBar:2008fao,BaBar:2015wkg,BaBar:2016wgb,BaBar:2012mrf}, and LHCb~\cite{LHCb:2011tcp,LHCb:2015svh,LHCb:2019hip,Langenbruch:2020quq,LHCb:2021lvy} Collaborations, etc.

In a large number of the successful tests of the SM, the experimental group has made good progress in the decay of $b\to s \ell^+\ell^-$, while different theoretical groups are also attracted by the interesting obervables in this decay channel. For example, as a parity violating observable, the longitudinal lepton polarization asymmetries of $B\to K\ell^+\ell^-$  can provide valuable information on the flavor changing loop effects in SM, which is discussed by QCD light-cone sum rule (LCSR) method~\cite{Aliev:1997gh} and the relativistic quark model based on the light-front formalism~\cite{Geng:1996az}. Among them, the results of branching ratio are consistent with experiments in terms of order of magnitude. In addition, as equally important observables, the dilepton invariant mass spectra and the forward-backward asymmetries of $B\to K^{(*)}\ell^+\ell^-$ that provide information on the short-distance contributions dominated by the top quark loops, are investigated by LCSR method in SM and supersymmetry model~\cite{Ali:1999mm}. And similar studies can also be found in Ref.~\cite{Aliev:1996hb,Colangelo:1995jv,Melikhov:1997wp,Chen:2001ri,Ali:2002jg,Parrott:2022smq}.

According to the current theoretical and experimental researches, the $B\to K^{(\ast)}\ell^+\ell^-$ can be regarded as one of the classical rare decay, offering substantial evidence that the decay $b\to s \ell^+\ell^-$ serves as a crucial tool for probing the detailed structure of the SM at the loop level. This hot spot has attracted our attention to the relevant decay processes. And during our investigation, we find an interesting point that the experimental collaborations have not studied the process of $B^0\to K_0^*(1430)\ell^+\ell^-$, which also induced by $b\to s$ transition, and where $K_0^*(1430)$ is a $P$-wave scalar meson, possessing properties that are fundamentally different from those of pseudoscalar meson $K$ and vector meson $K^\ast$. In order to make this kind of research more perfect, theorists have carried on the calculation for these decay processes. For instance, in Ref.~\cite{Aliev:2007rq}, they studied rare $B^0\to K_0^*(1430)\ell^+\ell^-$ decay by using TFFs in the framework of three-point QCD sum rules to calculate the branching fractions, with results in the range $10^{-7}\sim10^{-9}$ after considering short and long distance effects. In Ref.~\cite{Mahata:2022cxf}, Mahata investigated $B^0\to K_0^*(1430)\ell^+\ell^-$ decay in non-universal $Z$' models and using a model-independent way. They presented branching fractions, longitudinal lepton polarization asymmetries and LFU ratios, in which, they predicted the value of $R_{K_0^*}$ in the range 0.23 to 0.81. Recently in 2024~\cite{Khosravi:2024zaj}, Khosravi calculated TFFs of $B_{(s)}$ to light scalar mesons include $K_0^*(1430)$ by using LCSR method and further discussed the longitudinal lepton polarization asymmetries and branching fractions for $B^0\to K_0^*(1430)\ell^+\ell^-$ and $B^0\to K_0^*(1430) \nu\bar{\nu}$ in the range of $10^{-7}\sim10^{-8}$ and $10^{-6}$ order respectively. And other people have also investigated this process in LCSR~\cite{Wang:2010dp,Wang:2008da}, pQCD~\cite{Li:2008tk}, light-front quark model (LFQM)~\cite{Chen:2007na}. Therefore, we decide to focus on the rare decay $B^0\to K_0^*(1430)\ell^+\ell^-$ in this work, which can help us to further understand the FCNC decay $b\to s\ell^+\ell^-$ process from the aspect of light scalar mesons.

Similarly, the $b\to s \nu\bar{\nu}$ as a FCNC transitions, is also sensitive to SM. However, the photon exchange can significantly amplify the theoretical uncertainty caused by the breaking of the factorization. In this respect, the effect of photon exchange on the $b\to s\nu\bar{\nu}$ transitions is particularly small in comparison to that on the $b\to s\ell^+\ell^-$ transition~\cite{Buras:2014fpa}. Thus, some predictions of $b\to s \nu\bar{\nu}$ observables will be more accurate. The SM branching fraction of $B^+\to K^+\nu\bar{\nu}$ decay was determined by TFFs calculated by the lattice QCD, and its result is $\mathcal{B}(B^+\to K^+\nu\bar{\nu})=6.09(53)\times 10^{-7}$~\cite{Parrott:2022zte}. Recently in 2023, the Belle-II Collaboration reported the prediction of $\mathcal{B}(B^+\to K^+\nu\bar{\nu})$ to be $(2.3\pm 0.5^{+0.5}_{-0.4})\times 10^{-5}$ by using a $362~{\rm fb}^{-1}$ sample of electron-positron collisions at the $\Upsilon(4S)$ resonance collected with the Belle II detector at the SuperKEKB collider~\cite{Belle-II:2023esi}. There is some experimental data on $b\to s\nu\bar{\nu}$ from Belle~\cite{Belle:2017oht}, Belle-II~\cite{Dattola:2021cmw,Belle-II:2021rof}, and BaBar~\cite{BaBar:2013npw}, and theoretical data in Refs.~\cite{Altmannshofer:2009ma,Buras:2010pz,Buras:2012jb,Buras:2013qja}. In this regard, the $B^0\to K^\ast_0(1430)\nu\bar{\nu}$ decay has not yet been experimentally measured so far. Therefore, we can theoretically calculate it and look forward to combining it with future experimental results to help us better test the SM and determine more precise SM parameters. So next, we will also discuss the $B^0\to K_0^*(1430)\nu\bar{\nu}$ decay process in this paper.

In addition, as exclusive decay, the long-distance effects of $B^0\to K^\ast_0(1430)\ell^+\ell^-(\nu\bar{\nu})$ in the meson transition amplitudes of the effective Hamiltonian, are encoded in the TFFs. Therefore, in order to further calculate some observable quantities, we must first obtain the exact behaviors of TFFs for $B^0\to K_0^*(1430)$ decay within a reasonable $q^2$-range. In this regard, the TFFs of $B^0\to K_0^*(1430)$ decay have been calculated in different methods, such as LCSR~\cite{Han:2013zg,Sun:2010nv,Wang:2008da,Khosravi:2024zaj}, three-point sum rules (3pSR)~\cite{Yang:2005bv,Aliev:2007rq}, the pQCD~\cite{Li:2008tk} and LFQM~\cite{Chen:2007na}. Meanwhile, their respective numerical results of TFFs are also used to calculate the valuable observations of the $B^0\to K_0^*(1430)\ell^+\ell^-$. As far as current theoretical research is concerned, the LCSR method is suitable for calculating heavy to light TFFs, whose reliability has been widely recognized~\cite{Cheng:2017bzz,Fu:2018yin,Duplancic:2008ix,Fu:2020uzy,Ball:1998kk,Ball:2004rg,Bharucha:2015bzk,Wang:2015vgv,
Tian:2023vbh,Gao:2019lta,Khodjamirian:2011ub,Fu:2020vqd,Cheng:2018ouz,Fu:2014uea,Tian:2024lrn}. With the help of the quark-hadron duality and hadronic dispersion relation, the LCSR adopts the operator product expansion (OPE) near the light-cone $x^2\rightsquigarrow 0$ in terms of nonlocal operators, where the matrix elements of nonlocal operators within LCSR are parameterized as the final meson distribution amplitudes (DAs) arranged with different twist structures. In 2022, we have studied the TFFs for $B_s\to K_0^\ast(1430)$ through LCSR method, which is induced by $b\to u$ transition~\cite{Huang:2022xny}. In this work, based on LCSR, the TFFs $f^{B^0\to K_0^\ast}_{\pm,\rm{T}}$ will have similar analytic formulas to it, and the rationality of this prediction will be discussed in the next chapter. Moreover, it is worth noting that the internal structure of scalar mesons like $K_0^*(1430)$ is still controversial among physicists. In this case, two typical schemes for the classification of the scalar mesons are proposed in Refs.~\cite{Tornqvist:1982yv, Jaffe:1976ih, Kataev:2004ve, Vijande:2005jd, Jaffe:1976ig}, which provide a great and acceptable theoretical basis for physicists to study the scalar mesons. In scenario-1 (S1), the lighter scalar mesons $f_0(980),a_0(980),\kappa(800)$ below 1 GeV, etc, are seen as the lowest lying $q\bar{q}$ states. In scenario-2 (S2), the scalar mesons $f_0(1370),a_0(1450),K_0^*(1430)$ above 1 $\rm{GeV}$, etc, are treated as ground $q\bar{q}$ states. In this work, we assume the scalar meson $K_0^*(1430)$ is $q\bar{q}$ states.

In particular, in LCSR method, the $K_0^*(1430)$-meson twist-2 $\phi_{2;K_0^*}(x,\mu)$ and twist-3 DAs $\phi_{3;K_0^*}^p(x,\mu)$ and $\phi_{3;K_0^*}^\sigma(x,\mu)$ are the main sources of nonperturbative uncertainty for TFFs as a universal nonperturbative object, whose precise behavior is important for improving the prediction of the $B^0\to K_0^*(1430)$ decay. It plays an important role in describing the momentum fraction distributions of partons in $K_0^*(1430)$ in the lowest Fock state. Generally, the DAs of scalar mesons can be expanded in Gegenbauer series. Based on this, for twist-2 DA of $K_0^\ast(1430)$, one have also studied its behavior by different methods, such as LCSR~\cite{Sun:2010nv}, QCDSR~\cite{Cheng:2005nb}, and LF approach~\cite{Chen:2021oul}. Particularly, in order to avoid the unreliability of high-order Gegenbauer moments in QCDSR, a phenomenological light-cone harmonic oscillator (LCHO) model for $\phi_{2;K_0^*}(x,\mu)$ has been studied in our previous work~\cite{Huang:2022xny} on the basis of the Brodsky-Huang-Lepage (BHL) prescription~\cite{BHL}, which has been used to construct the leading twist DAs of different mesons~\cite{Hu:2023pdl,Zhong:2022ugk,Hu:2024tmc,Wu:2022qqx,Zhong:2022ecl,Hu:2021lkl,Hu:2021zmy,Zhong:2020cqr,Zhong:2018exo,Fu:2016yzx,
Zhong:2016kuv}. In which, the model parameters are determined by the least squares method. Meanwhile, for twist-3 DAs $\phi_{3;K_0^*}^p(x,\mu)$ and $\phi_{3;K_0^*}^\sigma(x,\mu)$, although we can choose a suitable chiral current to make its contribution very small, almost close to zero, in order to improve the accuracy of the TFFs behavior, we need to consider the contributions of twist-3 DAs. The twist-3 DAs of scalar mesons have been studied in Ref.~\cite{Lu:2006fr, Cheng:2005nb} whithin QCDSR. We have investigated the twist-3 DAs of scalar mesons within the framework of the QCD background field theory, whose details can also been found in our previous work~\cite{Han:2013zg}. Given the relatively limited experimental and theoretical attention devoted to the DAs of scalar mesons compared to other mesons, such as $\pi, K, \rho, K^\ast$-mesons~\cite{LatticeParton:2022zqc, Ball:2007zt, Hua:2020gnw, Chetyrkin:2007vm, Ball:2004ye}, and the positive feedback we have previously received from applying the LCHO model $\phi_{2;K_0^*}(x,\mu)$ to the $B_s, D_s\to K^\ast_0(1430)\ell\nu$ induced by $b\to u \ell\nu$ decay, this paper decides to further apply it to the specific decay process of $b\to s \ell^+\ell^-$. This exercise will serve as an opportunity to validate the feasibility of applying the LCHO model $\phi_{2;K_0^*}(x,\mu)$ to different decay processes, while adopting the universal truncation form of Gegenbauer polynomial series for twist-3 DAs. Their expressions will be briefly described in the next section.

This paper is organized as follows. In Section~\ref{Sec:II}, we present the effective Hamiltonian of the $b\to s \ell^+\ell^- ( \nu\bar{\nu})$. Then, we calculate the TFFs of $B^0\to K_0^*(1430)$ decay by the LCSR method, and briefly present the twist-2, 3 LCDAs. In Section~\ref{Sec:III}, we show our numerical results on TFFs and use it to calculate the differential decay widths, branching ratios, and lepton polarization asymmetries. Section~\ref{Sec:IV} is used to be a summary.

%
%

\section{Theoretical Framework}\label{Sec:II}

Effective Hamiltonian for the exclusive decay $B^0\to K_0^*(1430)\ell^+\ell^-$ at quark level induced by $b\to s\ell^+\ell^-$ transition can be described as~\cite{Grinstein:1988me}
\begin{align}
\mathcal{H}_{\rm eff}(b\to s\ell^+\ell^-) = \frac{4G_{\rm F} V_{tb}V_{ts}^\ast}{\sqrt{2}} \sum_{i=1}^{10} C_i(\mu){\cal O}_i(\mu),
\end{align}
where $G_{\rm F}$ is the Fermi constant, $V_{ij}$ are the CKM matrix elements and $C_i$ are Wilson coefficients at scale $\mu$ taken to be the $m_b$, and ${\cal O}_i$ are basis operators. These operators provide the main contribution in SM and the explicit forms for the decay $B\to K^\ast_0(1430)\ell^+\ell^-$ are
\begin{align}
&{\cal O}_1 = (\bar{s}_{\alpha} \gamma^{\mu} (1-\gamma_5) b_\alpha)(\bar{c}_\beta \gamma_{\mu}(1-\gamma_5) c_\beta),\nonumber\\
&{\cal O}_2 = (\bar{s}_{\alpha} \gamma^{\mu} (1-\gamma_5) b_\beta)(\bar{c}_\beta \gamma_{\mu}(1-\gamma_5) c_\alpha),\nonumber\\
&{\cal O}_7 =\frac{e}{16 {\pi^2}}m_b (\bar{s}_\alpha \sigma^{\mu \nu}(1+\gamma_5) b_{\alpha})F_{\mu \nu},\nonumber\\
&{\cal O}_9 =\frac{e^2}{16 {\pi^2}}(\bar{s}_\alpha \gamma^{\mu}(1-\gamma_5) b_\alpha)\bar{\ell}\gamma_{\mu}\ell ,\nonumber\\
&{\cal O}_{10} =\frac{e^2}{16 {\pi^2}} (\bar{s}_\alpha \gamma^{\mu} (1-\gamma_5)b_\alpha)\bar{\ell}\gamma_{\mu}\gamma_5 \ell.
\label{eq.O1,10}
\end{align}
Here, $F_{\mu\nu}$ is the electromagnetic field strength tensors, $\alpha$ and $\beta$ are colour indices, $\sigma^{\mu\nu}= \frac{i}{2}[\gamma^{\mu},\gamma^{\nu}]$. The coefficients of the QCD penguin operators ${\cal O}_{3-6}$ are small, so that its contribution can be neglected in SM. And the effective Hamiltonian for $b\to s\nu\bar{\nu}$ transition in the SM can be written as~\cite{Buras:2014fpa}
\begin{align}
\mathcal{H}_{\rm eff}(b\to s \nu\bar{\nu}) = -\frac{4G_F}{\sqrt{2}} V_{tb}V_{ts}^\ast C_L^{\rm SM}{\cal O}_L,
\label{Eq:Hamiltonian-bsvv}
\end{align}
where the Wilson coefficients are given by left handed current $C_L^{\rm SM}$ and ${\cal O}_L$ is four-fermion operator, which can be written as
\begin{align}
& C_L^{\rm SM} = -\frac{X(x_t)}{\sin^2\theta_w },\nonumber\\
& X(x_t)= \eta_X X_0(x_t),\nonumber\\
& {\cal O}_L = \frac{e^2}{16\pi^2} (\bar{s}\gamma_{\mu} (1-\gamma_5)b) (\bar{\nu}\gamma^{\mu}(1-\gamma_5)\nu),
\label{Eq:OL}
\end{align}
with $\eta_X\approx 1$ and $\theta_w$ is the Weinberg angle. The Inami-Lim function $X_0(x_t)$, which describes the short-distance $t$-quark contribution~\cite{Inami:1980fz}, will be given below. Here the effective Hamiltonian of Eq.~\eqref{Eq:Hamiltonian-bsvv} does not contain ${\cal O}_R$ due to the suppression of the right handed current $C_R^{\rm SM}$ by $m_s/m_d$, so we can ignore it safely~\cite{Bause:2023mfe}.

Then, the free quark amplitude for $b\to s\ell^+\ell^-$ and $b\to s\nu\bar{\nu}$ can be derived as~\cite{Chen:2007na}
\begin{align}
&\mathcal{M}(b\to s\ell^+\ell^-) = \langle s \ell^+\ell^-| \mathcal{H}_{\rm eff} |b \rangle
\nonumber\\
&\qquad = \frac{G_F \alpha_{\rm em}}{\sqrt{2}\pi} V_{ts}^\ast V_{tb} \Big[C_9^{\rm eff}(m_b) \left( \bar{s}\gamma_{\mu} (1-\gamma_5)b\bar{\ell}\gamma^{\mu}\ell \right)
\nonumber\\
&\qquad + C_{10}\left( \bar{s}\gamma_{\mu} (1-\gamma_5)b\bar{\ell}\gamma^{\mu}\gamma_5\ell \right)
\nonumber\\
&\qquad- 2iC_7^{\rm eff}(m_b) \frac{m_b }{q^2} \left(\bar{s}\sigma_{\mu\nu}q^{\nu}(1+\gamma_5)b\bar{\ell}\gamma^{\mu}\ell \right) \Big],
\label{Eq:Heff-bsll}
\\
&\mathcal{M}(b\to s\nu\bar{\nu}) = \langle s \nu\bar{\nu}| \mathcal{H}_{\rm eff} |b \rangle
\nonumber\\
&\qquad = \frac{G_F \alpha_{\rm em}}{2\pi\sqrt{2} \sin^2\theta_w} V_{ts}^\ast V_{tb}X_0(x_t)
\nonumber\\
&\qquad \times \bar{b}\gamma_{\mu}(1-\gamma_5)s\bar{\nu}_\ell \gamma_{\mu}(1-\gamma_5)\nu_\ell,
\label{Eq:Heff-bsvv}
\end{align}
where $\alpha_{\rm em}$ is the fine structure constant at $Z$-boson mass scale. The $C_{10}$ is independent on the energy scale, because there is no $Z$-boson in the effective theory, the operator ${\cal O}_{10}$ cannot be introduced by the insertion of four-quark operators~\cite{Wang:2008da}. In addition, the corresponding quark decay amplitude can receive short-distance and long-distance contributions from the matrix element of current-current operators ${\cal O}_1$ and ${\cal O}_2$, which can be quantified into the effective Wilson coefficient $C_9^{\rm eff}(m_b)$~\cite{Maji:2018gvz}. $C_9^{\rm eff}(m_b)= C_9+ Y(\hat{s})$, where $Y(\hat{s}) = Y_{\rm pert}(\hat{s}) +Y_{\rm LD}$ contains both the perturbative part $Y_{\rm pert}(\hat{s})$ and long-distance part $Y_{\rm LD}$ from the four quark operators~\cite{Buras:1994dj,Buchalla:1995vs}
\begin{align}
Y_{\rm pert}(\hat{s}) &= h(z,\hat{s})(3C_1 + C_2 + 3C_3 + C_4 + 3C_5 + C_6) \nonumber\\
& - \frac{1}{2}h(1,\hat{s})(4C_3 + 4C_4 + 3C_5 + C_6)\nonumber\\
& - \frac{1}{2}h(0,\hat{s})(C_3 + 3C_4) \nonumber\\
& + \frac{2}{9}(3C_3 + C_4 +3C_5 + C_6),
\label{Eq:Ypert}
\end{align}
with
\begin{align}
h(z,\hat{s})&= -\frac{8}{9}\ln z + \frac{8}{27} + \frac{4}{9}x - \frac{2}{9}(2+x)|1-x|^{1/2}\nonumber\\
&\times \begin{cases}
&\bigg|\dfrac{\sqrt{1-x} + 1}{\sqrt{1-x} - 1} \bigg| - i\pi, x\equiv 4z^2/\hat{s} < 1
\\
&2\arctan\dfrac{1}{\sqrt{x-1}}, x\equiv 4z^2/\hat{s} > 1
\end{cases}
\end{align}
and the $h(0,\hat{s})$ have the expression
\begin{align}
h(0,\hat{s})=\frac{8}{27} - \frac{8}{9}\ln \frac{m_b}{\mu} - \frac{4}{9}\ln\hat{s} + \frac{4}{9}i\pi,
\end{align}
where $z= m_c/m_b$ and $\hat{s}=q^2/m_b^2$.

The exclusive decay processes $B^0\to K_0^*(1430)\ell^+\ell^-$ and $B^0\to K_0^*(1430)\nu\bar{\nu}$ are described in terms of the matrix elements of the quark operators in Eqs.~\eqref{Eq:Heff-bsll} and \eqref{Eq:Heff-bsvv} over meson states. Thus, one should essentially to know the $B^0\to K_0^*(1430)$ transition matrix elements with forms of $\langle K_0^*(1430)|\bar{s}\gamma_{\mu}\gamma_5 b|B^0\rangle$ and $\langle K_0^*(1430)|\bar{s}\sigma_{\mu\nu}\gamma_5 q^{\mu} b|B^0\rangle$, which can be parameterized in terms of TFFs, {\it i.e.,}
\begin{align}
&\langle K_0^*(1430)(p)|\bar{s} i \gamma_\mu \gamma_5 b|B^0(p+q)\rangle
\nonumber\\
&\qquad\qquad = 2 p_\mu f^{B^0\to K_0^\ast}_+(q^2) + q_\mu \tilde f^{B^0\to K_0^\ast}(q^2),
\nonumber\\
&\langle K_0^*(1430)(p)|\bar{s} \sigma_{\mu\nu} \gamma_5 q^\nu b|B^0(p+q)\rangle
\nonumber\\
&\qquad\qquad = -[2p_\mu q^2 - 2q_\mu(p\cdot q)]\frac{ f^{B^0\to K_0^\ast}_{\rm T}(q^2)}{m_{B^0}+m_{K_0^*}},
\label{Eq:MatrixElement}
\end{align}
where $\tilde f^{B^0\to K_0^\ast}_+(q^2) = [f^{B^0\to K_0^\ast}_+(q^2) + f^{B^0\to K_0^\ast}_-(q^2)]$. Then we can get the general expression of differential decay rates for $B^0\to K_0^*(1430)\ell^+\ell^-(\nu\bar{\nu})$ by using Eqs.~\eqref{Eq:Heff-bsll}, \eqref{Eq:Heff-bsvv} and \eqref{Eq:MatrixElement}, respectively, which are expressed as~\cite{Sun:2010nv,Khosravi:2022fzo}
\begin{align}
&\frac{d\Gamma(B^0\to K_0^*(1430)\ell^+\ell^-)}{dq^2}  = \frac{G_F^2 |V_{tb} V_{ts}^\ast|^2 m_{B^0}^3 \alpha_{\rm em}^2}{1536\pi^5}
\nonumber\\
& \times \sqrt{1\!-\!\frac{4r_{\ell}}{s}}\left[\left(1\!+\!\frac{2r_{\ell}}{s}\right)\varphi_S^{3/2} \alpha_S(q^2) + \varphi_S^{1/2} r_{\ell} \delta_S(q^2) \right],\label{eq:DWll}
\end{align}
where the abbreviations have the definitions $s=q^2/m_{B_0
}^2$, $r_\ell=m_{\ell}^2/m_{B^0}^2$, $r_S=m_{K_0^*}^2/m_{B^0}^2$, $\varphi_S=(1-r_S)^2- 2s(1+r_S)+ s^2$. The helicity form factors $\alpha_S(q^2)$, $\delta_S(q^2)$ have the relationship with traditional TFFs, which have the following expressions,
\begin{align}
\alpha_S(q^2)&=\big|C_9^{\rm eff}f^{B^0\to K_0^\ast}_+(q^2) - \frac{2}{1+\sqrt{r_S}} C_7^{\rm eff}f^{B^0\to K_0^\ast}_{\rm T}(q^2)\big|^2
\nonumber\\
& + \big|C_{10}f^{B^0\to K_0^\ast}_+(q^2)\big|^2,
\label{eq.alphas}
\\
\delta_S(q^2)&=6|C_{10}|^2 \bigg\{[2(1+r_S)-s] \big|f^{B^0\to K_0^\ast}_+(q^2)\big|^2
\nonumber\\
& + (1-r_S) 2{\rm Re}[f^{B^0\to K_0^\ast}_+(q^2)f^{B^0\to K_0^\ast}_-(q^2)]
\nonumber\\
& + s\big|f^{B^0\to K_0^\ast}_-(q^2)\big|^2 \bigg\},
\label{eq.deltas}
\end{align}
and
\begin{align}
&\frac{d\Gamma (B^0 \to K_0^*(1430) \nu\bar{\nu})}{dq^2} =\frac{G_F^2 |V_{tb} V_{ts}^\ast|^2 m_{B_0
}^3 \alpha_{\rm em}^2}{256\pi^5}
\nonumber\\
&\qquad\qquad\qquad \times \frac{|X_0(x_t)|^2}{\sin^4 \theta_w}\varphi_S^{3/2} |f^{B^0\to K_0^\ast}_+(q^2)|^2,
\label{eq:DWvv}
\end{align}
where $x_t = m_t^2/m_W^2$ and
\begin{align}
X_0(x_t)=\frac{x_t}{8}\bigg[\frac{2+x_t}{x_t-1} + \frac{3x_t-6}{(x_t-1)^2}~\ln x_t\bigg].
\end{align}

Then, we need to obtain the behavior of the TFFs for $B^0\to K_0^*(1430)$ by LCSR. We first follow the standard steps of sum rules to introduce the vacuum-to-$K_0^*(1430)$ correlation function, which is related to the TFFs $f^{B^0\to K_0^\ast}_\pm(q^2)$ and $f^{B^0\to K_0^\ast}_{\rm T}(q^2)$, which is defined as
\begin{align}
\Pi_\mu(p,q) &= i\int d^4x e^{iq\cdot x} \langle K_0^*(p)|{\rm T}\{ j_{2\mu}(x), j_1(0)\}|0\rangle \nonumber\\
&= F(q^2,(p+q)^2)p_\mu + \tilde{F}(p^2,(p+q)^2) q_\mu,
\nonumber\\
\tilde{\Pi}_\mu(p,q) &= i\int d^4x e^{iq\cdot x} \langle K_0^*(p)| {\rm T}\{\tilde{j}_{2\mu}(x), j_1(0)\}|0\rangle \nonumber\\
&= F^{\rm T}(q^2,(p+q)^2) [p_\mu q^2-q_\mu(p \cdot q)].
\label{eq:correlatorTFFs}
\end{align}
where the currents have the form that $j_{2\mu}(x)= \bar{q}_2(x)\gamma_\mu \gamma_5 b(x)$, $\tilde{j}_{2\mu}(x)= \bar{q}_2(x) \sigma_{\mu\nu} \gamma_5 q^\nu b(x)$ and $j_1(0)=m_b\bar{b}(0) i\gamma_5 q_1(0)$. Meanwhile, the symbols $q_{1}$ and $q_{2}$ denote the light $d$ and $s$-quark, respectively. And $p$ is the $K_0^*(1430)$-meson four momentum, $q$ and $(p+q)$ are transition momentum and the $B^0$-meson momentum, respectively. According to the basic steps of LCSR method, we can derive the hadronic representation by inserting a complete intermediate state with $B^0$-meson quantum numbers into the hadron current of Eq.~\eqref{eq:correlatorTFFs} in the time-like $q^2$-region and separate the pole term of the lowest $B^0$-meson.
\begin{align}
\Pi_\mu^{\rm{H}}(p,q)&=\frac{\langle K_0^*|\bar{s}\gamma_\mu \gamma_5 b|B^0\rangle \langle B^0| \bar{b} i\gamma_5 d|0\rangle m_b}{m_{B^0}^2 - (p+q)^2}\nonumber\\
&+\sum_{\rm{H}}\frac{\langle K_0^*|\bar{s}\gamma_\mu \gamma_5 b|B^{0\rm{H}}\rangle \langle B^{0\rm{H}}| \bar{b} i\gamma_5 d|0\rangle m_b}{m_{B^{0\rm{H}}}^2 - (p+q)^2},\nonumber\\
\tilde{\Pi}_\mu^{\rm{H}}(p,q)&=\frac{\langle K_0^*|\bar{s} \sigma_{\mu\nu} \gamma_5 q^\nu b|B^0\rangle \langle B^0| \bar{b} i\gamma_5 d|0\rangle m_b}{m_{B^0}^2 - (p+q)^2}\nonumber\\
&+\sum_{\rm{H}}\frac{\langle K_0^*|\bar{s} \sigma_{\mu\nu} \gamma_5 q^\nu b|B^{0\rm{H}}\rangle \langle B^{0\rm{H}}| \bar{b} i\gamma_5 d|0\rangle m_b}{m_{B^{0\rm{H}}}^2 - (p+q)^2}.
\end{align}
In which, corresponding vacuum-meson matrix element of $B^0$-meson decay constant $\langle B^0| \bar{b}i\gamma_5 d|0\rangle m_b = m^2_{B^0} f_{B^0}$ and Eq.~\eqref{Eq:MatrixElement} will also be used. After replacing the contributions of higher reaonances and continuum states with dispersion relation, the invariant amplitudes $F^{\rm{H}}, \tilde{F}^{\rm{H}}$ and $F^{\rm{H}}_{\rm T}$ can read as
\begin{align}
&F^{\rm{H}}(p^2,(p+q)^2)=\frac{-2i m_{B^0}^2 f_{B^0} f^{B^0\to K_0^\ast}_+(q^2)}{ m_{B^0}^2 - (p+q)^2}
\nonumber\\
& \qquad +\int^{s_0^{B^0}}_{m_b^2}ds \frac{\rho_+^{\alpha_s}(s)}{s-(p+q)^2}+\int^{\infty}_{s_0^{B^0}}ds \frac{\rho_+(s)}{s-(p+q)^2},
\nonumber\\
&\tilde F^{\rm{H}}(p^2,(p+q)^2) =\frac{-i m_{B^0}^2 f_{B^0} [f^{B^0\to K_0^\ast}_+(q^2) + f^{B^0\to K_0^\ast}_-(q^2)]}{ m_{B^0}^2 - (p+q)^2}
\nonumber\\
& \qquad +\int^{s_0^{B^0}}_{m_b^2}ds \frac{\rho_{\pm}^{\alpha_s}(s)}{s-(p+q)^2}+\int^{\infty}_{s_0^{B^0}}ds \frac{\rho_{\pm}(s)}{s-(p+q)^2},
\nonumber\\
&F^{\rm{H}}_{\rm T}(p^2,(p+q)^2) =\frac{-2m_{B^0}^2\! f_{B^0} f^{B^0\to K_0^\ast}_{\rm T}(q^2)}{(m_{B^0} + m_{K_0^*}) [m_{B^0}^2- (p+q)^2]}
\nonumber\\
& \qquad +\int^{s_0^{B^0}}_{m_b^2}ds \frac{\rho_{\rm{T}}^{\alpha_s}(s)}{s-(p+q)^2}+\int^{\infty}_{s_0^{B^0}}ds\frac{\rho_{\rm{T}}(s)}{s-(p+q)^2}.
\end{align}
This hadronic representation is similar to our pervious study of $B_s\to K_0^\ast(1430)$~\cite{Huang:2022xny}, where the input parameters variations are $m_{B_s}\to m_{B^0}$ and $f_{B_s}\to f_{B^0}$. Moreover, another difference between this work and our previous study is that the TFFs consider the NLO corrections to the twist-2 terms from Ref.~\cite{Wang:2014vra}. In simpler terms, by utilizing standard dimensional regularization and the $\overline{\rm MS}$ scheme, they calculated six Feynman diagrams that contribute to the twist-2 terms with perturbative $\mathcal{O}(\alpha_s)$ corrections, obtaining the relevant renormalized correlation functions. Subsequently, using dispersion relations, they derived the corresponding QCD spectral densities $\rho_{\pm,\rm{T}}^{\alpha_s}$ in the region $s\in[m_b^2,s^{B^0}_0]~\rm{GeV}$. The specific derivation process has been discussed in Ref.~\cite{Wang:2014vra}, and for convenience, we present their final expressions in Appendix~\ref{Appendix} Then, after performing OPE near the light-cone $x^2\rightsquigarrow 0$ in the deep Euclidean region $q^2=-Q^2\ll0$, and with the help of quark hadron duality and Borel transform, the similar TFFs formulas can be easily obtained,
\begin{widetext}
\begin{eqnarray}
f_+^{B^0\to K_0^\ast}(q^2) &=&\frac{m_b \bar{f}_{K_0^*} e^{m_{B^0}^2/M^2}}{2m_{B^0}^2 f_{B^0}} \int^{\tilde{u}_0}_{u_0} du e^{-(m_b^2-\bar{u}q^2+u\bar{u}m_{K_0^*}^2)/(uM^2)}\bigg\{-m_b \frac{\phi_{2;K_0^*}(u,\mu)}{u}
\nonumber\\
&+&  m_{K_0^*} \phi_{3;K_0^*}^p(u,\mu)+  m_{K_0^*}\bigg[\frac{2}{u}+\frac{4u m_b^2 m_{K_0^*}^2}{(m_b-q^2+u^2 m_{K_0^*}^2)^2}- \frac{m_b^2+q^2-u^2 m_{K_0^*}^2}{m_b^2-q^2+u^2 m_{K_0^*}^2}
\nonumber\\
&\times& \frac{d}{du}\bigg] \frac{\phi_{3;K_0^*}^\sigma(u,\mu)}{6}\bigg\}+\frac{ \bar{f}_{K_0^*} e^{m_{B^0}^2/M^2}}{2m^2_{B^0}f_{B^0}} \int^{s^{B^0}_0}_{m_b^2} ds \rho_+^{\alpha_s}(s)e^{-\frac{s}{M^2}}
\label{eq:LCSRTFFs1}
\\
\tilde f^{B^0\to K_0^\ast}(q^2) &=& \frac{m_b \bar{f}_{K_0^*} m_{K_0^*} e^{m_{B_0}^2/M^2}}{m_{B^0}^2 f_{B^0}}\int^{\tilde{u}_0}_{u_0} du e^{-(m_b^2-\bar{u}q^2+u\bar{u}m_{K_0^*}^2)/(uM^2)}\bigg[ \frac{\phi_{3;K_0^*}^p(u)}{u}
\nonumber\\
&+& \frac{1}{u} \frac{d}{du} \frac{\phi_{3;K_0^*}^\sigma(u,\mu)}{6} \bigg] + \frac{\bar{f}_{K_0^*} e^{m_{B^0}^2/M^2}}{m_{B^0}^2 f_{B^0}} \int^{s^{B^0}_0}_{m_b^2} ds \rho_{\pm}^{\alpha_s}(s)e^{-\frac{s}{M^2}},
\label{eq:LCSRTFFs2}
\\
f_{\rm T}^{B^0\to K_0^\ast}(q^2) &=& \frac{(m_{B^0} + m_{K_0^*}) m_b \bar{f}_{K_0^*} e^{m_{B^0}^2/M^2}}{m_{B^0}^2 f_{B^0}}
\int^{\tilde{u}_0}_{u_0} du e^{-(m_b^2-\bar{u}q^2+u\bar{u}m_{K_0^*}^2)/(uM^2)}
\nonumber\\
&\times& \Bigg\{ -\frac{\phi_{2;K_0^*}(u,\mu)}{2u} + \frac{m_b m_{K_0^*}}{m_b^2-q^2+u^2m_{K_0^*}^2}  \left[ \frac{2um_{K_0^*}^2}{m_b^2-q^2+u^2m_{K_0^*}^2} - \frac{d}{du} \right]
\nonumber\\
&\times& \frac{\phi_{3;K_0^*}^\sigma(u,\mu)}{6} \Bigg\} +\frac{(m_{B^0} + m_{K_0^*}) \bar{f}_{K_0^*} e^{m_{B^0}^2/M^2}}{2 m_{B^0}^2 f_{B^0}} \int^{s^{B^0}_0}_{m_b^2} ds \rho_{\rm T}^{\alpha_s}(s)e^{-\frac{s}{M^2}}.
\label{eq:LCSRTFFs3}
\end{eqnarray}
\end{widetext}
where
\begin{align}
u_0 & = \Big[ \sqrt{(q^2 - s^{B^0}_0 + m_{K_0^*}^2)^2 + 4m_{K_0^*}^2 (m_b^2 - q^2)}
\nonumber\\
& + q^2 - s^{B^0}_0 + m_{K_0^*}^2 \Big]/(2m_{K_0^*}^2),
\\
\tilde{u}_0 &= \Big[ \sqrt{(q^2 - m_b^2 + m_{K_0^*}^2)^2 + 4m_{K_0^*}^2 (m_b^2 - q^2)}
\nonumber\\
&+ q^2 - m_b^2 + m_{K_0^*}^2 \Big]/(2m_{K_0^*}^2).
\end{align}

Next, we will briefly introduce the leading twist-2, 3 DAs from our previous work~\cite{Huang:2022xny,Han:2013zg}. The twist-2 DA can be constructed by twist-2 wave function (WF), which can be obtained by using LCHO model. Based on BHL description, the equal-times WF in the rest frame have a connection with light-cone WF in the infinite momentum frame~\cite{BHL}. And then, by adopting the spin-space WF $\chi_{2;K_0^\ast}(x,\mathbf{k}_\perp)$ from Wigner-Melosh rotation and integrating over the transverse momentum dependence, the twist-2 $\phi_{2;K_0^*}(x,\mu)$ DA can be written as
\begin{align}
&\phi_{2;K_0^*}(x,\mu) = \frac{A_{2;K_0^*} \beta_{2;K_0^*} \tilde{m}}{4\sqrt{2}\pi^{3/2}} \sqrt{x\bar{x}} \varphi_{2;K_0^*}(x)
\nonumber\\
&\qquad\times \exp \left[-\frac{\hat{m}_q^2 x + \hat{m}_s^2 \bar{x} - \tilde{m}^2}{8\beta^2_{2;K_0^*} x\bar{x}} \right]
\nonumber\\
&\qquad\times \left\{{\rm Erf} \left( \sqrt{\frac{\tilde{m}^2+\mu^2}{8\beta^2_{2;K_0^*}x\bar{x}}}\right)-{\rm Erf}\left(\sqrt{\frac{\tilde{m}^2}{8\beta^2_{2;K_0^*}x\bar{x}}}\right)\right\},
\label{eq:leading-twist}
\end{align}
where Erf$(x) = 2\int_0^x e^{-t^2}dx/\sqrt{\pi}$ is the error function, $\hat{m}_s = 280~{\rm MeV}, \hat{m}_q =250~{\rm MeV}$ are the light constituent quark mass with $\tilde{m}=\hat{m}_q x + \hat{m}_s \bar{x}$. The longitudinal distribution function $\varphi_{2;K_0^*}(x) = (x\bar{x})^{\alpha_{2;K_0^*}} \Big[ C_1^{3/2}(2x - 1)+\hat{B}_{2;K_0^*}C_2^{3/2}(2x - 1) \Big]$. These above unknown model parameters $A_{2;K_0^*}, \alpha_{2;K_0^*}, \beta_{2;K_0^*}$ can be determined by using the least squares method to fit the first ten $\xi$-moment of $K_0^*(1430)$ twist-2 LCDA. The detailed calculation process can be known from the Ref.~\cite{Zhong:2022ecl,Zhong:2021epq}.

In addition, the twist-3 LCDAs of scalar mesons can be generally expanded into a series of Gegenbauer polynomials and taken the truncated form to remain the first few terms,
\begin{align}
\phi_{3;K_0^*}^p(x,\mu) & = 1+a_{1,p}^{3;K_0^*}(\mu) C_1^{1/2}(\xi) + a_{2,p}^{3;K_0^*}(\mu) C_2^{1/2}(\xi),
\nonumber\\
\phi_{3;K_0^*}^\sigma(x,\mu) & = 6x\bar{x} \Big[ 1+ a_{1,\sigma}^{3;K_0^*}(\mu) C_1^{3/2}(\xi) + a_{2,\sigma}^{3;K_0^*}(\mu) C_2^{3/2}(\xi)\Big],
\label{eq:twist-3,sigma}
\end{align}
where $\xi=(2x-1)$, the Gengenbauer moments $a_{1,p}^{3;K_0^*}(\mu)= 0.0108\pm 0.0015, a_{2,p}^{3;K_0^*}(\mu)= 0.1342\pm 0.0172, a_{1,\sigma}^{3;K_0^*}(\mu)= 0.0231\pm 0.0085$ and $a_{2,\sigma}^{3;K_0^*}(\mu)= 0.0145\pm 0.0028$ at $\mu= 3~\rm{GeV}$. There exists a certain relationship between the Gegenbauer moments and the $\langle \xi^n_{p/\sigma;K_0^\ast} \rangle$ moments on the same scale. The corresponding $\xi$-moment has been calculated by using the background field method in QCD sum rule~\cite{Han:2013zg}.

\section{Numerical Analysis}\label{Sec:III}

In order to calculate the TFFs of the rare $B^0\to K_0^*(1430) \ell^+\ell^- / \nu\bar{\nu}$ decays, we adopt following input parameters: the heavy meson mass $m_{B^0}=5279.66\pm 0.12~\rm{MeV}$ and $m_{K_0^*}=1425\pm 50~\rm{MeV}$, the $b$ and $d$-quark $\overline{\rm MS}$ masses $m_b(\bar{m}_b)=4.18_{-0.03}^{+0.04}~\rm{GeV}$ and $m_d(2~\rm{GeV}) = 4.67_{-0.17}^{+0.48}~\rm{MeV}$~\cite{ParticleDataGroup:2024cfk}, decay constants $f_{B^0}=0.207^{+0.017}_{-0.009}~\rm{GeV}$~\cite{Gelhausen:2013wia} and $f_{K_0^*}=0.427\pm 0.085~\rm{GeV}$ at $\mu=1~\rm{GeV}$~\cite{Du:2004ki}. For the typical process energy scale of the $B^0\to K_0^*(1430)$ transition, we take $\mu_k=\sqrt{m_{B^0}^2-m_b^2}\backsimeq 3~\rm{GeV}$ in this work. For the effective threshold parameters and the Borel windows, we adopt $s^{B^0}_0=36\pm 1~\rm{GeV}$ and $M^2 = 25\pm1~\rm{GeV^2}$, respectively. Meanwhile, for the first ten $\xi$-moment of twist-2 DA, we have already determined $s_0$ based on the criterion that 0-order $\xi$-moment equals 1, which is the continuum threshold parameter for $\langle \xi^n_{2;K_0^\ast} \rangle$, subsequently determined Borel paramete $M^2$ for $\langle \xi^n_{2;K_0^\ast} \rangle$ by setting the condition of continum state and dimension-six terms for odd and even moment, and finally obtained the results for $\langle \xi^n_{2;K_0^\ast} \rangle$. Then, the optimal results for the model parameters $A_{2;K_0^*}, \alpha_{2;K_0^*}, \beta_{2;K_0^*}$ can be determined through the least squares method at $\mu_k=3~\rm{GeV}$~\cite{Huang:2022xny},
\begin{align}
A_{2;K_0^\ast} &= -67~\rm{GeV^{-1}},\nonumber\\
\alpha_{2;K_0^\ast} &= -0.148,\nonumber\\
\beta_{2;K_0^*}& = 1.088~\rm{GeV}.
\label{eq:optimal results}
\end{align}
\begin{table*}[t]
\renewcommand{\arraystretch}{1.5}
\begin{center}
\footnotesize
\caption{Numerical results for the $B^0\to K_0^*(1430)$ TFFs at large recoil region within errors. As a comparison, the theoretical predictions are also listed.}
\label{table:TFFsvalue}
\begin{tabular}{l l l l}
\hline
References~~~~~~~~~~~~~~~ &$\hh f^{B^0\to K_0^*}_+(0)$~~~~~~~~~~~~~ &$\hh f^{B^0\to K_0^*}_-(0)$~~~~~~~~~~~ &$\hh f^{B^0\to K_0^*}_{\rm T}(0)$\\ \hline
This work                &$\hh 0.470_{-0.101}^{+0.086}$               &$-0.340_{-0.068}^{+0.068}$                &$\hh 0.537_{-0.115}^{+0.112}$\\
3pSR~\cite{Aliev:2007rq} &$\hh 0.31_{-0.08}^{+0.08}$               &$-0.31_{-0.07}^{+0.07}$                &$-0.26_{-0.07}^{+0.07}$\\
pQCD(S1)~\cite{Li:2008tk} &$-0.34_{-0.09}^{+0.07}$                  &$\hh /$                                &$-0.44_{-0.11}^{+0.10}$\\
pQCD(S2)~\cite{Li:2008tk} &$\hh 0.60_{-0.15}^{+0.18}$               &$\hh /$                                &$\hh 0.78_{-0.19}^{+0.25}$\\
LFQM~\cite{Chen:2007na}   &$-0.26$                                  &$\hh 0.21$                             &$\hh 0.34$\\
LCSR~\cite{Han:2013zg}    &$\hh 0.45_{-0.05}^{+0.06}$               &$-0.28_{-0.06}^{+0.06}$                &$\hh 0.46_{-0.05}^{+0.06}$\\
LCSR~\cite{Sun:2010nv}    &$\hh 0.49$                               &$-0.49$                                &$\hh 0.69$\\
LCSR~\cite{Han:2023pgf}   &$\hh 0.43$                               &$-0.42$                                &$\hh 0.58$\\
LCSR~\cite{Wang:2014upa}  &$\hh 0.523_{-0.07}^{+0.07}$              &$-0.275_{-0.064}^{+0.064}$             &$\hh 0.657_{-0.109}^{+0.109}$\\
\hline
\end{tabular}
\end{center}
\end{table*}

\begin{table*}[t]
\renewcommand{\arraystretch}{1.5}
\begin{center}
\footnotesize
\caption{The extrapolation parameters $\beta_{0,1,2}$ and quality of fit $\Delta$ for the $B^0\to K_0^*(1430)$ TFFs $f_{\pm,\rm{T}}^{{B^0\to K_0^*}} (q^2)$ based on the SSE approach.}
\label{table:parameters}
\begin{tabular}{lllll}
\hline
~~~~~~~~~~~~~~~~~~~~~~~&~~~~~~~~~~~~~&$\hh f_+^{{B^0\to K_0^*}}(q^2)$~~~~~~~~~&$\hh f_-^{{B^0\to K_0^*}}(q^2)$~~~~~~~~~~ &$\hh f_{\rm T}^{{B^0\to K_0^*}}(q^2)$\\ \hline
                       & $\beta_0$   &$\hh 0.556$                             &$-0.270$                                  &$\hh 0.647$\\
                       & $\beta_1$   &$-1.600$                                &$\hh 0.385$                                  &$-1.288$\\
\raisebox{2ex}[0pt]{upper limits}    & $\beta_2$   &$\hh 1.819$               &$\hh 4.575$                               &$\hh 4.296$\\
                       & $\Delta$    &$\hh 0.007\%$                           &$\hh 0.070\%$                             &$\hh 0.015\%$\\ \hline
                       & $\beta_0$   &$\hh 0.469$                              &$-0.338$                                   &$\hh 0.535$\\
                       & $\beta_1$   &$-1.363$                                 &$\hh 0.428 $                                  &$-1.075$\\
\raisebox{2ex}[0pt]{central value}   & $\beta_2$   &$\hh 1.462$                &$\hh 4.995$                                &$\hh 3.352$\\
                       & $\Delta$    &$\hh 0.006\%$                           &$\hh 0.064\%$                             &$\hh 0.014\%$\\ \hline
                       & $\beta_0$   &$\hh 0.368 $                             &$-0.406$                                   &$\hh 0.421$\\
                       & $\beta_1$   &$-1.089$                                 &$\hh 0.513$                                   &$-0.866$\\
\raisebox{2ex}[0pt]{lower limits}     & $\beta_2$   &$\hh 1.060$                &$\hh 5.946$                               &$\hh 2.409$\\
                       & $\Delta$    &$\hh 0.007\%$                           &$\hh 0.064\%$                             &$\hh 0.013\%$\\ \hline
\end{tabular}
\end{center}
\end{table*}

With the above input parameters and the $K_0^*(1430)$-meson DA parameters, we can obtain the values of the TFFs at zero momentum transfer, $i.e.$, $q^2 = 0~{\rm GeV^2}$ and the errors caused by different input parameters are as follow
\begin{align}
f_+^{B^0\to K_0^\ast}(0)&=0.452+\left(_{-0.013}^{+0.011}\right)_{s_0^{B^0}}+\left(_{-0.002}^{+0.002}\right)_{M^2}
\nonumber\\
&+\left( _{-0.009}^{+0.006}\right)_{m_b} \!+\! \left(_{-0.009}^{+0.009}\right)_{m_{B^0},m_{K_0^\ast}} \!+\! \left(_{-0.089}^{+0.090}\right)_{f_{K_0^\ast}}
\nonumber\\
& + \left( _{-0.034}^{+0.021}\right)_{f_{B^0}}+\left(\pm 0.018\right)_{\rm{NLO}}
\nonumber\\
&=0.470_{-0.101}^{+0.086},
\\
f_-^{B^0\to K_0^\ast}(0)&=-0.306+\left(_{-0.024}^{+0.023}\right)_{s_0^{B^0}}+\left(_{-0.002}^{+0.002}\right)_{M^2} \nonumber\\
&+\left(_{-0.013}^{+0.014}\right)_{m_b} \!+\! \left(_{-0.014}^{+0.014}\right)_{m_{B^0},m_{K_0^\ast}} \!+\! \left(_{-0.119}^{+0.118}\right)_{f_{K_0^\ast}}
\nonumber\\
&+\left( _{-0.032}^{+0.041}\right)_{f_{B^0}} +\left(\pm 0.034\right)_{\rm{NLO}}
\nonumber\\
&=-0.340_{-0.068}^{+0.068},
\\
f_{\rm{T}}^{B^0\to K_0^\ast}(0)& =0.522+\left(_{-0.009}^{+0.007}\right)_{s_0^{B^0}} + \left(_{-0.003}^{+0.003}\right)_{M^2} 
\nonumber\\
&+\left(_{-0.010}^{+0.007}\right)_{m_b} \!+\! \left(_{-0.012}^{+0.012}\right)_{m_{B^0},m_{K_0^\ast}} \!+\! \left(_{-0.104}^{+0.103}\right)_{f_{K_0^\ast}}
\nonumber\\
&+\left( _{-0.039}^{+0.024}\right)_{f_{B^0}} + \left(\pm 0.015\right)_{\rm{NLO}}
\nonumber\\
&=0.537^{+0.112}_{-0.115}.
\end{align}
The perturbation $\mathcal{O}(\alpha_s)$ corrections to the TFFs $f_+^{B^0\to K_0^\ast}(0), f_-^{B^0\to K_0^\ast}(0)$ and $f_{\rm{T}}^{B^0\to K_0^\ast}(0)$ are $4\%$, $10\%$ and $3\%$, respectively. In the mean time, we also include the predictions for TFFs $f_{\pm,\rm{T}}^{B^0\to K_0^\ast}(0)$ of the other approaches to compare with our results in Table~\ref{table:TFFsvalue}, such as the 3pSR method~\cite{Aliev:2007rq}, pQCD~\cite{Li:2008tk},  LFQM method~\cite{Chen:2007na}, and LCSR method~\cite{Han:2013zg,Sun:2010nv,Han:2023pgf,Wang:2014upa}. As can be seen from the Table~\ref{table:TFFsvalue}, the result of $f^{B^0\to K_0^\ast}_+(0)$ is in good agreement with the center value from LCSR~\cite{Wang:2014upa,Han:2013zg,Han:2023pgf,Sun:2010nv} within a reasonable error range. And compared with 3pSR~\cite{Aliev:2007rq}, pQCD~\cite{Li:2008tk} in different schemes and LFQM~\cite{Chen:2007na}, there are great differences in $f^{B^0\to K_0^\ast}_+(0)$, mainly because of the different methods. In addition, our numerical results for $f^{B^0\to K_0^\ast}_-(0)$ and $f^{B^0\to K_0^\ast}_{\rm T}(0)$ agree well with 3pSR~\cite{Aliev:2007rq} and LCSR~\cite{Han:2023pgf}, respectively.
\begin{figure*}[t]
\begin{center}
\includegraphics[width=0.42\textwidth]{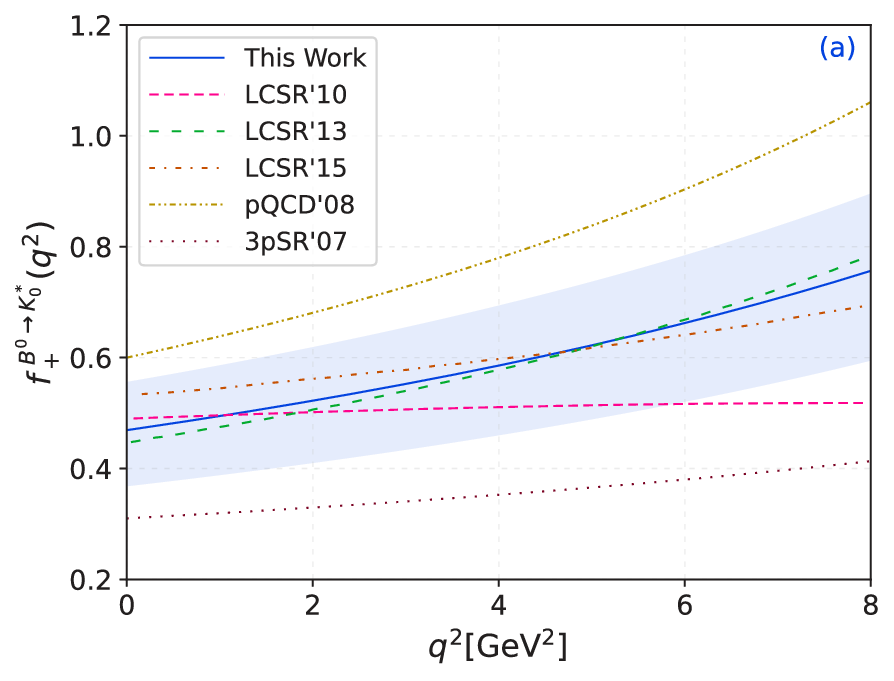}\includegraphics[width=0.42\textwidth]{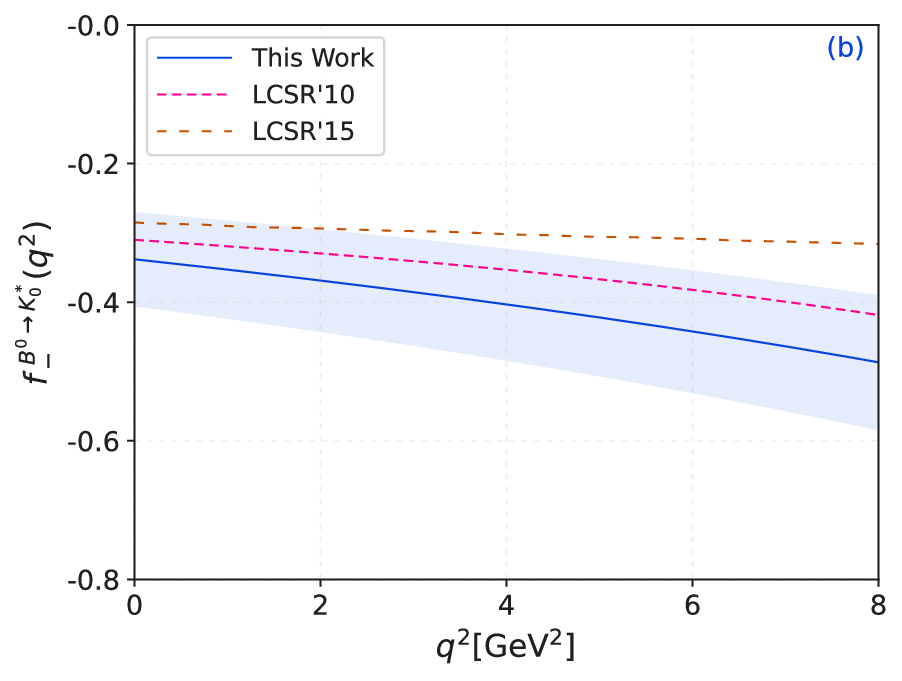}
\includegraphics[width=0.42\textwidth]{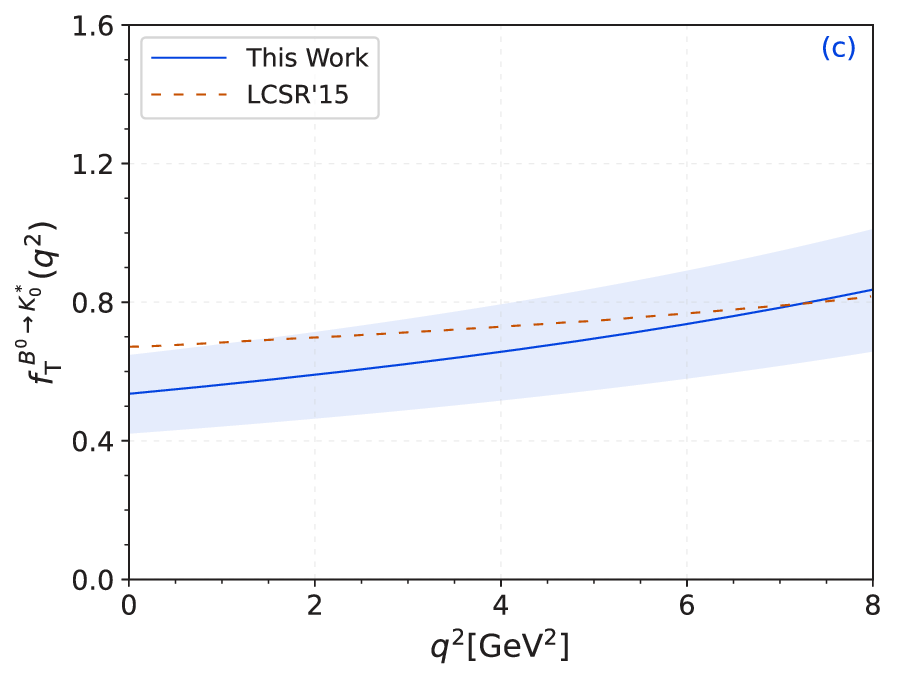}\includegraphics[width=0.42\textwidth]{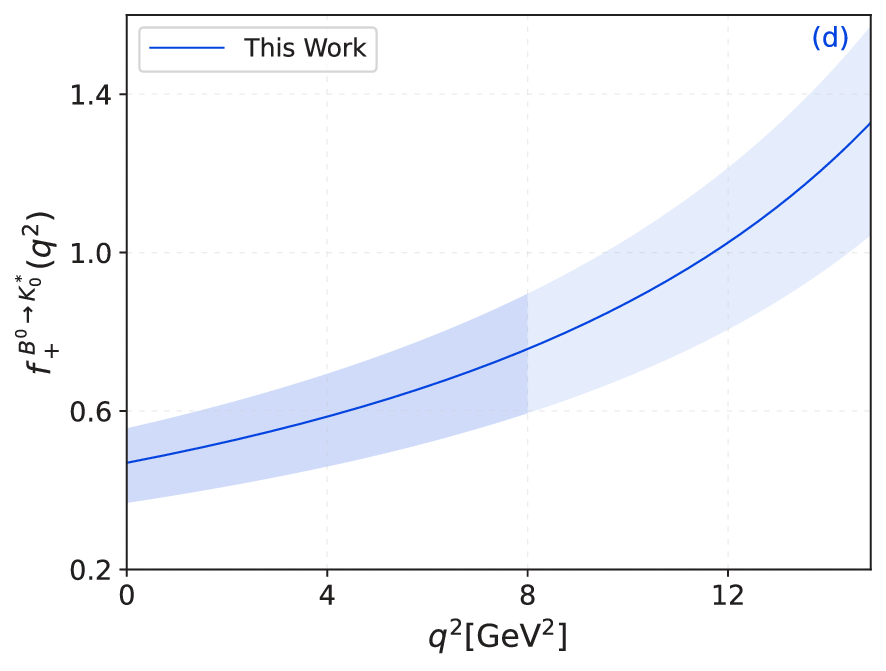}\\
\end{center}
\caption{Behaviors of the TFFs $f_{\pm,\rm{T}}^{B^0\to K_0^\ast}$ in the low and intermediate $q^2$-region $i.e.$, planes ($a$), ($b$), ($c$) and the $f_+^{B^0\to K_0^\ast}(q^2)$ in whole $q^2$-region, $i.e.$, plane ($d$). The solid line represents the central value and shaded bands corresponds to uncertainties. The lines in different colors represent theoretical predictions from 3pSR 2007~\cite{Aliev:2007rq},  pQCD 2008~\cite{Li:2008tk}, LCSR 2010~\cite{Sun:2010nv}, LCSR 2013~\cite{Han:2013zg} and LCSR 2015~\cite{Wang:2014vra}. Their TFFs behaviors are used to compare with our results.}
\label{fig:TFFs}
\end{figure*}
The physically allowable range for the TFFs is $0< q^2 < (m_{B^0} - m_{K_0^*})^2 \thickapprox 14.85~\rm{GeV^2}$, but the LCSR approach for $B^0\to K_0^*(1430)$ TFFs are only reliable in low and intermediate region,  $i.e.$, $q^2 \in [0,8]~\rm{GeV^2}$. In order to calculate the observables in phenomenology, such as branching fractions and lepton polarization asymmetries of the $B^0\to K_0^*(1430)\ell^+\ell^- (\nu\bar{\nu})$ decays, we need extrapolate the TFFs in whole kinematical region $0< q^2 < (m_{B^0} - m_{K_0^*})^2$ via $z(q^2,t_0)$ converging the simplified series expansion (SSE)~\cite{Bharucha:2010im,Bharucha:2015bzk}. The TFFs take the following form

\begin{align}
f^{B^0 \to K_0^*}_i(q^2) &= \frac{1}{1-q^2/m_{B^0}}\sum_{k=0,1,2}{\beta_kz^k(q^2,t_0)},
\label{eq:fi}
\end{align}
where $f^{B^0 \to K_0^*}_i(q^2)$ with $i=(+,-,{\rm T})$ represent the three TFFs $f^{B^0 \to K_0^*}_{\pm}(q^2)$ and $f^{B^0 \to K_0^*}_{\rm{T}}(q^2)$, respectively. $\beta_k$ are real coefficients and $z(q^2,t)$ is the function,
\begin{align}
z^k(q^2,t_0 ) =\frac{\sqrt{t_+-q^2}-\sqrt{t_+-t_0}}{\sqrt{t_+-q^2}+\sqrt{t_+-t_0}},
\label{eq:zk}
\end{align}
with $t_{\pm} = (m_{B^0} \pm m_{K_0^*})^2$ and $t_0=t_+(1-\sqrt{1-t_-/t_+})$. The most important is that we need to fit the appropriate value of three free parameters $\beta_{0,1,2}$ to make the quality of extrapolation $\Delta$ as small as possible. Then the quality of extrapolation $\Delta$ is defined as
\begin{align}
\Delta =\frac{\sum_t{|F_i(t)-F_i^{\rm fit}|}}{\sum_t{|F_i(t)|}}~\times 100.
\label{eq:Delta}
\end{align}
\begin{figure*}[t]
\centering
\includegraphics[width=0.42\textwidth]{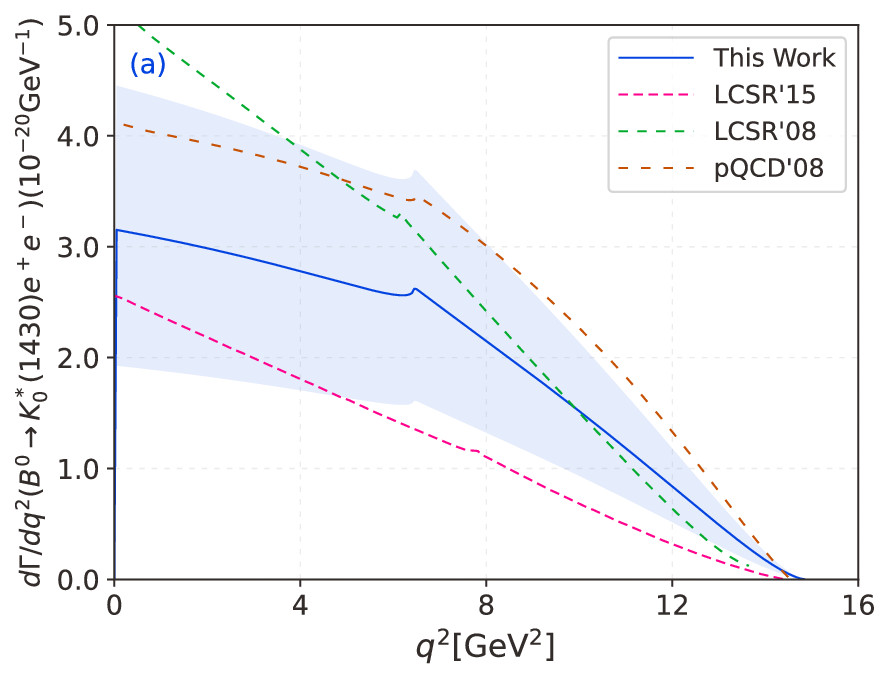}\includegraphics[width=0.42\textwidth]{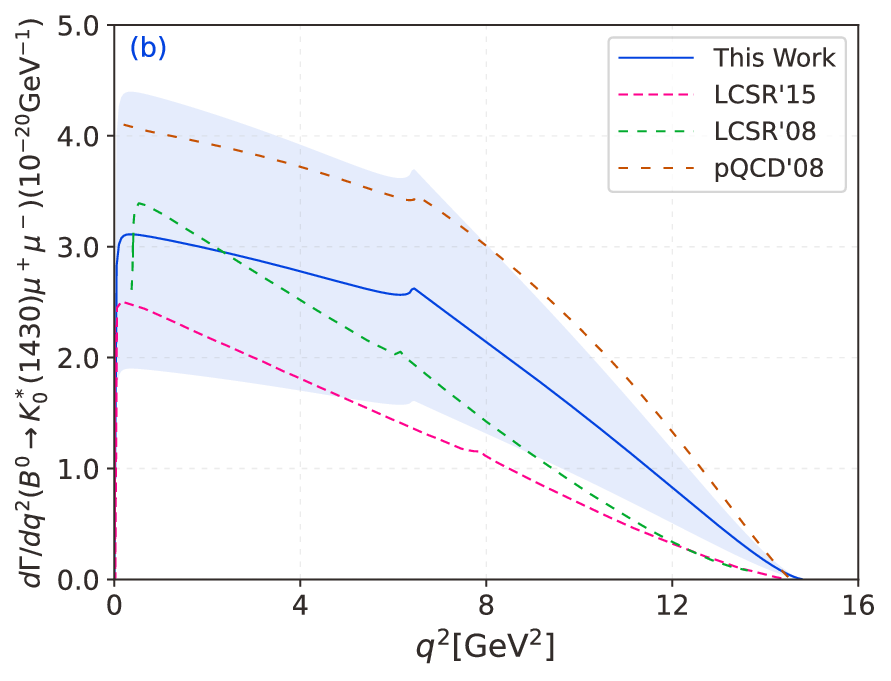}
\includegraphics[width=0.42\textwidth]{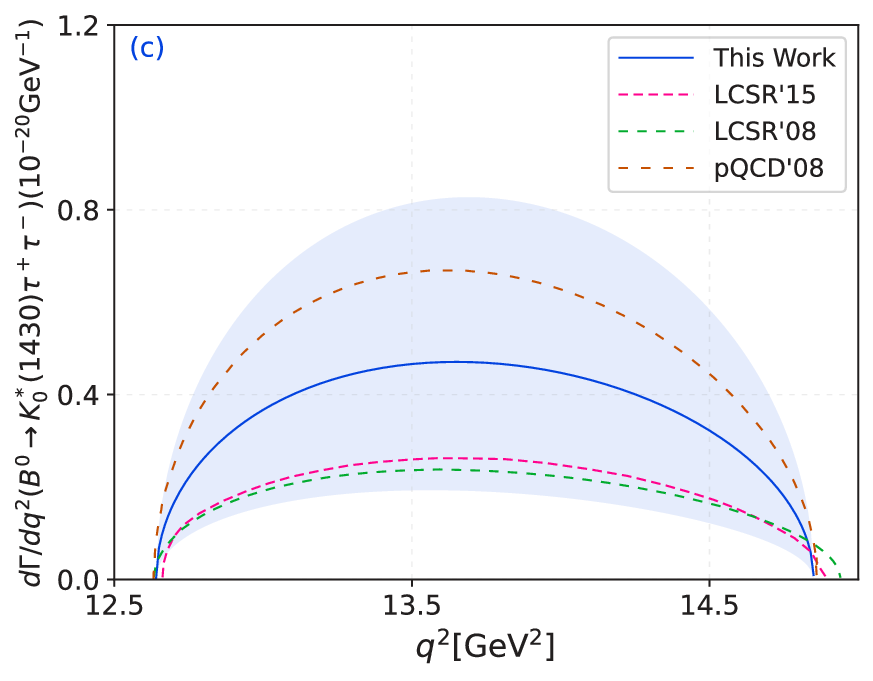}\includegraphics[width=0.42\textwidth]{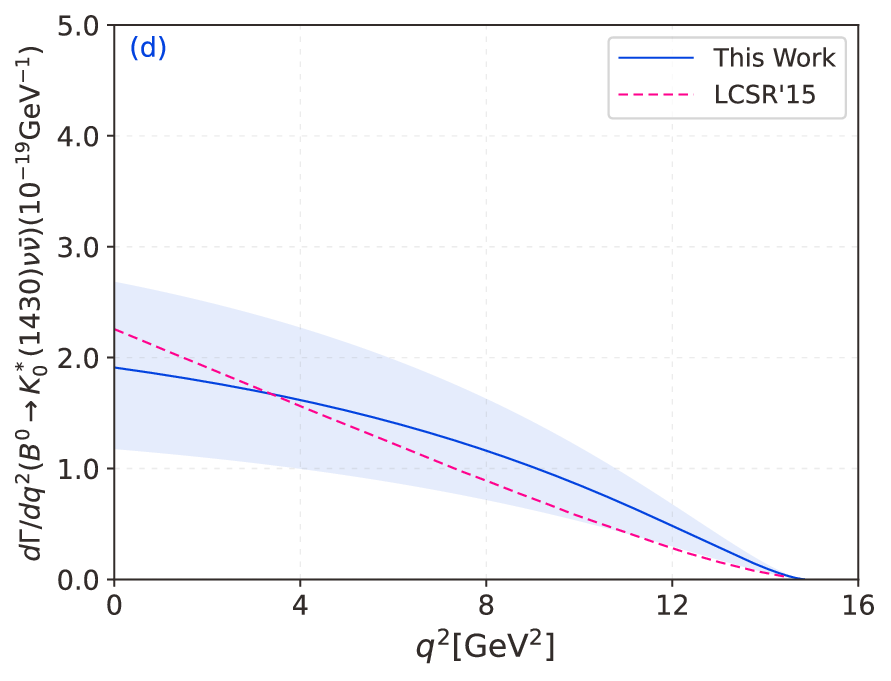}\\
\caption{The differential decay widths as functions of $q^2$, where the (\colorr{a}), (\colorr{b}), (\colorr{c}) and (\colorr{d}) denote the decay $B^0\to K_0^\ast(1430)e^+e^-$, $B^0\to K_0^\ast(1430)\mu^+\mu^-$, $B^0\to K_0^\ast(1430)\tau^+\tau^-$ and $B^0\to K_0^\ast(1430)\nu\bar{\nu}$. The other results from LCSR 2015~\cite{Wang:2014upa}, LCSR 2008~\cite{Wang:2008da} and pQCD~\cite{Li:2008tk} are used to compare with our results. }
\label{fig:differenal DW}
\end{figure*}
After extrapolating the TFFs $f_i^{B^0 \to K_0^*} (q^2)$ to the whole physical $q^2$-region, we can obtain the coefficients $\beta_{0,1,2}$ and $\Delta$, which are listed in Table~\ref{table:parameters}. We can observe that all the $\Delta$ values of $B^0\to K_0^\ast$ are no more than $0.070\%$, which indicate the effect of the extrapolation is perfect. The corresponding data in the Table~\ref{table:parameters} can help us to reproduce the results directly  of TFFs by Eqs.~(\ref{eq:fi}), (\ref{eq:zk}) and (\ref{eq:Delta}). The behavior of the $B^0\to K_0^*$ TFFs $f^{B^0 \to K_0^*}_{\pm,\rm{T}}(q^2)$ in low and intermediate $q^2$-region and the $f^{B^0 \to K_0^*}_+(q^2)$ in whole $q^2$-region are shown in Fig.~\ref{fig:TFFs}, respectively. Since these literatures 3pSR 2007~\cite{Aliev:2007rq}, pQCD 2008~\cite{Li:2008tk}, LCSR 2010~\cite{Sun:2010nv}, LCSR 2013~\cite{Han:2013zg}, and LCSR 2015~\cite{Wang:2014vra} depicted the behavior diagrams of the TFFs in the low and intermediate range of $0 < q^2 < 8 \rm{GeV^2}$, for the sake of convenience, we hereby present their predictions within this region and subsequently compare them with our outcomes, $e.g.$, Fig.~\ref{fig:TFFs}(\colorr{a}), (\colorr{b}), (\colorr{c}). The shadow band represents the error range, which is generated by the upper and lower limits of the input parameters. The solid line is our central result and other lines are the central predictions of other groups. Then, we can see that for $f^{B^0 \to K_0^*}_+ (q^2)$, our result is in a good agreement with the results of LCSR 2013~\cite{Han:2013zg} and LCSR 2015~\cite{Wang:2014vra} within reasonable error range. With the prediction of LCSR 2010~\cite{Sun:2010nv}, the agreement is only maintained probably in $q^2 \in[0,5.8]~\rm{GeV^2}$ and for the results of 3pSR 2007~\cite{Aliev:2007rq} and pQCD 2008~\cite{Li:2008tk}, there are significant differences compared to ours, which are due to the differences in the calculation methods. It is worth mentioning that in LCSR 2013~\cite{Han:2013zg}, the major contribution to the TFFs comes from twist-3 DAs by introducing proper chiral currents. For $f^{B^0 \to K_0^*}_- (q^2)$, our prediction agrees with QCD 2007~\cite{Aliev:2007rq} and when $q^2 > 2~\rm{GeV^2}$, the prediction of LCSR 2015~\cite{Wang:2014vra} is beyond our error range. For $f^{B^0\to K_0^\ast}_{\rm T}(q^2)$, the curve shows good consistency with the result of LCSR 2015~\cite{Wang:2014vra}. In addition, we present the behavior of $f^{B^0 \to K_0^*}_+ (q^2)$ in the whole $q^2$-region, $i.e.$, Fig. \ref{fig:TFFs}(\colorr{d}), where darker band represents the our prediction and the lighter band is the SSE prediction. We are more concerned about the extrapolation effect of $f^{B^0 \to K_0^*}_+ (q^2)$ than $f^{B^0 \to K_0^*}_{-,{\rm T}} (q^2)$, because in following calculations of observations, it makes a major contribution.

Next step, we can get the differential decay widths of $B^0\to K_0^*(1430) \ell^+\ell^-$ and $B^0\to K_0^*(1430) \nu\bar{\nu}$ by Eqs.~(\ref{eq:DWll}) and (\ref{eq:DWvv}), respectively. Before performing the calculation, we need to define some parameters. Such as $m_\mu= 105.658~{\rm MeV}$, $m_e=0.511~{\rm MeV}$, $m_{\tau}= 1776.82~{\rm MeV}$, $m_t=167~{\rm GeV}$, $m_W=80.4~{\rm GeV}$, fermi coupling constant $G_{\rm{F}}=1.166\times 10^{-5}~{\rm GeV^{-5}}$, CKM matrix elements $|V_{tb}|=0.9991, |V_{ts}|= 41.61\times 10^{-3}$, $\alpha_{\rm em}=1/137$ and $\sin^2\theta_w=0.23$. In addition, $C_7^{\rm eff}=-0.313, C_{10}=-4.669$~\cite{Grinstein:1988me}.

\begin{table*}[t]
\renewcommand{\arraystretch}{1.5}
\footnotesize
\begin{center}
\caption{The branching ratio values of the $B^0\to K_0^*(1430)\ell^+\ell^-(\ell=e,\mu,\tau)$ and $B^0\to K_0^*(1430)\nu\bar{\nu}$ within errors. Meanwhile, other theoretical predictions are also given as a comparison.}
\label{table:Br}
\begin{tabular}{lll}
\hline
~~~~~~~~~~~~~~~~~~~~~~~~~~~~~~~~~~~~~~~~~~&${\cal B}(B^0\to K_0^*(1430)e^+e^-)$~~~~~~~~~~~~~~~~ &${\cal B}(B^0\to K_0^*(1430)\mu^+\mu^-)$ \\ \hline
This work    &$6.65_{-2.42}^{+2.52}\times 10^{-7}$    &$6.62_{-2.41}^{+2.51}\times 10^{-7}$      \\
QCDSR~\cite{Aliev:2007rq} &$(2.09-2.68)\times 10^{-7}$     &$(2.07-2.66)\times 10^{-7}$        \\
LCSR~\cite{Wang:2010dp}   &$4.51\times 10^{-7}$            &$4.48\times 10^{-7}$          \\
LCSR~\cite{Wang:2008da}   &$5.7_{-2.4}^{+3.4}\times 10^{-7}$     &$5.6_{-2.3}^{+3.1}\times 10^{-7}$          \\
pQCD(S1)~\cite{Li:2008tk} &$3.13_{-1.21}^{+1.73}\times 10^{-7}$  &$3.13_{-1.21}^{+1.73}\times 10^{-7}$       \\
pQCD(S2)~\cite{Li:2008tk} &$9.78_{-4.40}^{+7.66}\times 10^{-7}$  &$9.78_{-4.40}^{+7.66}\times 10^{-7}$       \\
LFQM~\cite{Chen:2007na}    &$1.63\times 10^{-7}$          &$1.62\times 10^{-7}$      \\
LCSR~\cite{Wang:2014upa}  &$4.14\pm1.17\pm0.18\times 10^{-7}$    &$4.12\pm1.17\pm0.18\times 10^{-7}$       \\
\hline
            &${\cal B}(B^0\to K_0^*(1430)\tau^+\tau^-)$~~ &${\cal B}(B^0\to K_0^*(1430)\nu\bar{\nu})$ \\ \hline
This work   &$1.88_{-0.97}^{+1.10}\times 10^{-8}$         &$3.85_{-1.48}^{+1.55}\times 10^{-6}$\\
QCDSR~\cite{Aliev:2007rq}  &$(1.70-2.20)\times 10^{-9}$   &$--$\\
LCSR~\cite{Wang:2010dp}   &$7.35\times 10^{-9}$  &$--$\\
LCSR~\cite{Wang:2008da}   &$9.8_{-5.5}^{+12.4}\times 10^{-9}$  &$--$\\
pQCD(S1)~\cite{Li:2008tk} &$2.00_{-0.77}^{+1.16}\times 10^{-9}$  &$--$\\
pQCD(S2)~\cite{Li:2008tk} &$6.29_{-2.95}^{+5.71}\times 10^{-9}$  &$--$\\
LFQM~\cite{Chen:2007na}   &$2.86\times 10^{-9}$      &$1.16\times 10^{-6}$\\
LCSR~\cite{Wang:2014upa}  &$1.10\pm0.03\pm0.04\times 10^{-8}$  &$3.49\pm0.93\pm0.15\times 10^{-6}$\\\hline
\end{tabular}
\end{center}
\end{table*}

With these parameters, the differential decay widths with variations in $q^2$ can be obtained after substituting the derived $B^0\to K_0^*(1430)$ TFFs into the decay width formula, which is presented in Fig.~\ref{fig:differenal DW}. The solid line represents the central values. As a comparison, we also give the results of the  LCSR 2008~\cite{Wang:2008da}, pQCD~\cite{Li:2008tk}, and LCSR 2015~\cite{Wang:2014upa}. As shown in Fig.~\ref{fig:differenal DW}, there exists minor discontinuities in the differential decay width curves for electron and muon pair final states in the range of $6~{\rm GeV^2}< q^2 <8~{\rm GeV^2}$ in these results of references and our predictions, but this is not the case in the decay width of tauon pair channel. This is mainly because the $h(z,\hat{s})$ and $h(1,\hat{s})$ functions produce the discontinuities, but for the tauon pair channel, its $q^2$ minimum value is $4 m^2_{\tau} \simeq 12.6~{\rm GeV^2}$ , which is sufficiently large to guarantee that changes in $q^2$ do not pass through discontinuity points in the $h(z,\hat{s})$ and $h(1,\hat{s})$ functions. For the electron and muon pair channel, the trend of the curve is almost the same, with only minor differences observed in the large recoil region. Secondly, for electron and muon pair channel in Fig.~\ref{fig:differenal DW}(\colorr{a}) and \ref{fig:differenal DW}(\colorr{b}), there is different level of agreement with other results in different region. Since the TFF $f_+^{B^0\to K_0^{*}}(q^2)$ of pQCD~\cite{Li:2008tk} is larger than ours in entire $q^2$ region, its central value of the decay width is almost equal to our upper limit. From Fig.~\ref{fig:differenal DW}(\colorr{c}), we can roughly estimate from the ordinate that the maximum total decay width for tauon pair channel is an order of magnitude smaller than that of the electron and muon pair channels, primarily because of the heavily suppressed phase space. Our result is in good agreement with LCSR 2008~\cite{Wang:2008da}, LCSR 2015~\cite{Wang:2014upa}, and pQCD~\cite{Li:2008tk} within errors. Finally, for $B^0\to K_0^*(1430) \nu\bar{\nu}$ differential decay width, our result is in agreement with LCSR 2015~\cite{Wang:2014upa} in whole $q^2$-region. The Fig.~\ref{fig:differenal DW}(\colorr{a})-(\colorr{d}) show the results are converge to zero at the small recoil region $q^2_{\rm max} = (m_{B^0}-m_{K_0^\ast})^2$, which indicates that our predictions are reasonable.

After integrating over $q^2$ in the whole physical region $4m_\ell^2 \leq q^2 \leq (m_{B_0}-m_{K_0^*})^2$ for $B^0\to K_0^*(1430) \ell^+\ell^-$ and $0 \leq q^2 \leq(m_{B_0}-m_{K_0^*})^2$ for $B^0\to K_0^*(1430) \nu\bar{\nu}$, respectively, we can obtain the total decay widths,
\begin{align}
&\Gamma (B^0\to K_0^* e^+e^-)= 2.88_{-1.10}^{+1.17}\times 10^{-4}~\rm{GeV},\nonumber\\
&\Gamma (B^0\to K_0^* \mu^+\mu^-)= 2.94_{-1.10}^{+1.17}\times 10^{-4}~\rm{GeV},\nonumber\\
&\Gamma (B^0\to K_0^* \tau^+\tau^-)= 8.14_{-4.18}^{+4.75}\times 10^{-6}~\rm{GeV},\nonumber\\
&\Gamma (B^0\to K_0^* \nu\bar{\nu})= 1.67_{-0.64}^{+0.67}\times 10^{-3}~\rm{GeV}.
\label{value:decay width}
\end{align}
\begin{figure*}[t]
\begin{center}
\includegraphics[width=0.42\textwidth]{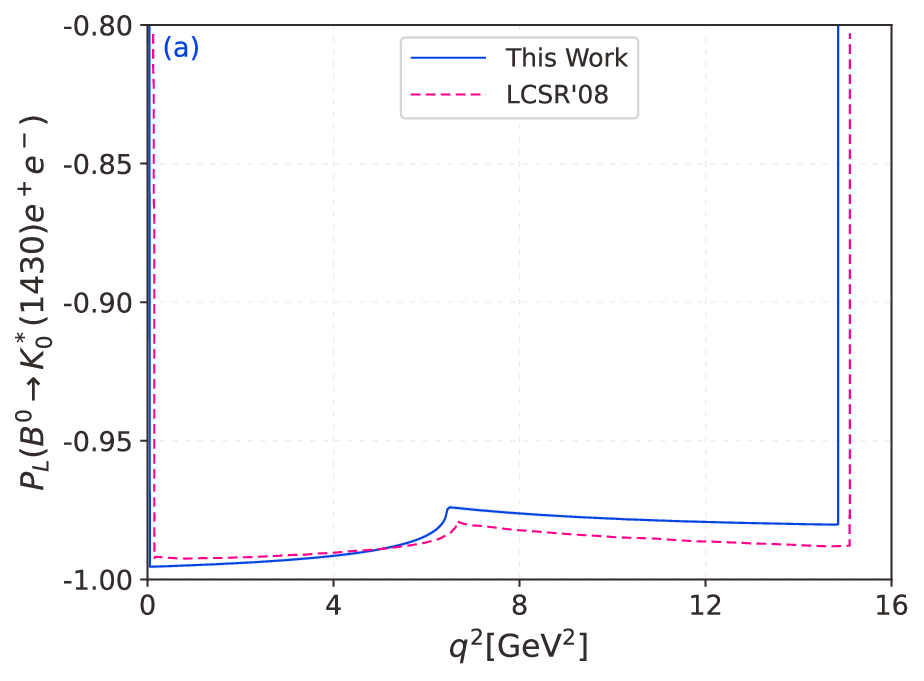}\includegraphics[width=0.42\textwidth]{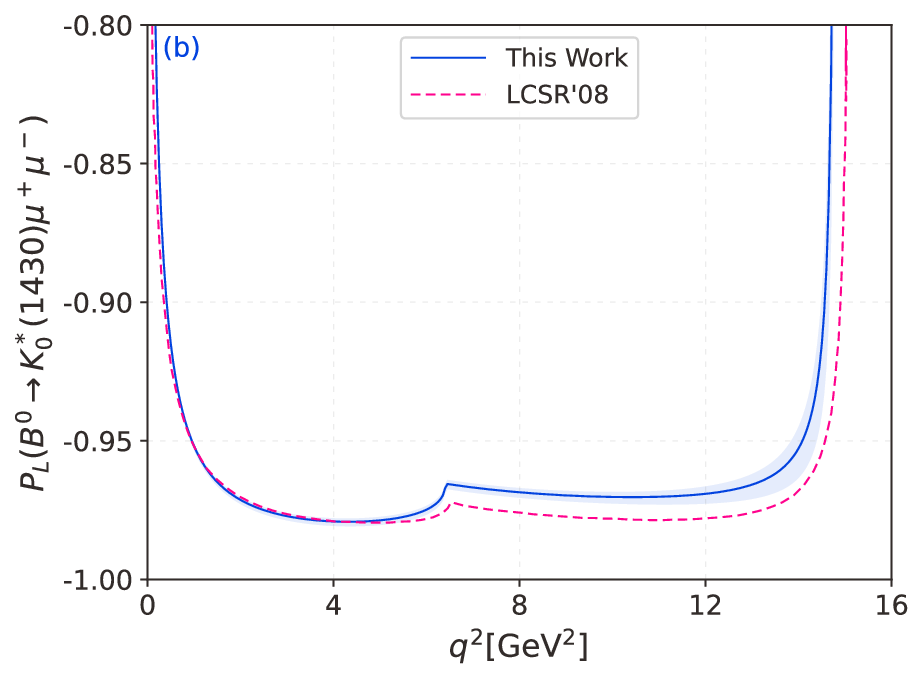}
\includegraphics[width=0.42\textwidth]{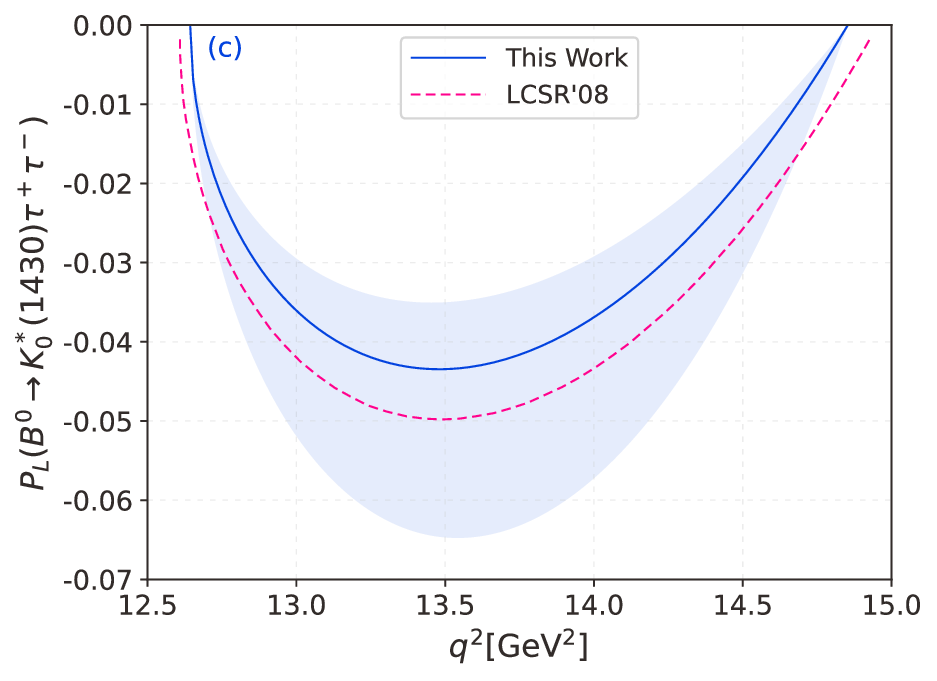}
\end{center}
\caption{Longitudinal lepton polarization asymmetries with variations of the $q^2$, where (\colorr a), (\colorr b) and (\colorr c) denote the decay $B^0\to K_0^\ast(1430)e^+e^-$, $B^0\to K_0^\ast(1430)\mu^+\mu^-$ and $B^0\to K_0^\ast(1430)\tau^+\tau^-$. Similarly, the graphic results of the LCSR 2008~\cite{Wang:2008da} are used to compare with our results.}
\label{fig:PL}
\end{figure*}

Next, by using the lifetime of the initial state $B^0$-meson, $\tau_{B^0}=1.519\times 10^{-12}$s, the branching ratios of the different decay channels $B^0\to K_0^* \ell^+\ell^-$ and $B^0\to K_0^* \nu\bar{\nu}$ can be obtained, which are shown in Table~\ref{table:Br}. In which, Table~\ref{table:Br} also displays the other branching fractions that are calculated in Refs.~\cite{Wang:2014upa, Li:2008tk, Wang:2008da, Wang:2010dp, Chen:2007na, Aliev:2007rq} for comparison. It is clear that our predictions for branching ratios of electron and muon pair final states should be basically the same in numerical results from total decay widths, and the branching ratio of tauon-pair channel is quite small because of the heavily suppressed phase space. Considering the efficiency of detecting tauon, it is still very challenging to experimentally measure the branching ratio for $B^0\to K_0^*(1430)\tau^+\tau^-$ compared to electron and muon pair channels. As shown in the Table~\ref{table:Br}, we can see that the branching fractions of the decay $B^0\to K_0^*(1430)e^+e^-/\mu^+\mu^-$ are of the order $10^{-7}$, which is in a good agreement with Ref.~\cite{Wang:2014upa, Li:2008tk, Wang:2008da, Wang:2010dp, Chen:2007na, Aliev:2007rq}. In particular, the center value of LCSR~\cite{Wang:2014upa,Wang:2008da,Wang:2010dp} is within our error range. For the branching fractions of the decays $B^0\to K_0^*(1430)\tau^+\tau^-$ and $B^0\to K_0^*(1430)\nu\bar{\nu}$ are of the order $10^{-8}$ and $10^{-6}$ respectively, which is in good agreement with the results from LCSR~\cite{Wang:2014upa} within error. In thses decay processes, the calculation of the TFFs $f^{B^0\to K_0^\ast}_+$ which accounts for the main contribution, can easily affect the value of the branching fraction. While there are also reasons for different input parameters. Compared to decay $B^0\to K_0^*(1430)\nu\bar{\nu}$, the decays $B^0\to K_0^*(1430)\ell^+\ell^-$ have a smaller branching ratio due to a smaller phase spaces, while these two decay processes are optimal in searching for new physics beyond SM.

Furthermore, due to the value of the forward-backward asymmetries for $B^0\to K_0^*(1430)\ell^+\ell^-$ are equal to zero in the SM~\cite{Geng:1996az,Belanger:1992dx}, we decide to only discuss the longitudinal lepton polarization asymmetries in this work, whose unit vector $\hat{e}_L = \vec{p}_\ell /|\vec{p}_\ell|= \pm 1$. Then the longitudinal lepton polarization asymmetries $P_L$ can be defined as
\begin{align}
P_L(s)&=  \dfrac{ \dfrac{d\Gamma(\hat{e}_L \hat{\xi} = 1)}{ds} - \dfrac{d\Gamma(\hat{e}_L \hat{\xi} = -1)}{ds}} {\dfrac{d\Gamma(\hat{e}_L \hat{\xi} = 1)}{ds} + \dfrac{d\Gamma(\hat{e}_L \hat{\xi} = -1)}{ds}},
\end{align}
Finally, we can obtain that~\cite{Wang:2008da}
\begin{align}
P_L(s)&= \frac{2(1-\frac{4r_\ell}{s})^{1/2}}{(1+\frac{2r_\ell}{s})\alpha_S\varphi_S + r_l\delta_S} {\rm Re} \Bigg[\varphi_S \Bigg(C_9^{\rm eff} \frac{f^{B^0\to K_0^\ast}_+(q^2)}{2}
\nonumber\\
& - 2\frac{C_7 f^{B^0\to K_0^\ast}_{\rm T}(q^2)}{1+ \sqrt{r_S}} \Big)\Big(C_{10}\frac{f^{B^0\to K_0^\ast}_+(q^2)}{2}\Bigg)^\ast \Bigg].
\label{eq:PL}
\end{align}

By substituting the relevant parameters into Eq.~\eqref{eq:PL}, we can obtain the longitudinal lepton polarization asymmetries as functions of squared momentum transfer $q^2$ for $B^0\to K_0^*(1430)$ with $q^2_{\rm min}= 4m_\ell^2$ and $q^2_{\rm max}=(m_{B^0}-m_{K_0^*})^2$, which are present in Fig.~\ref{fig:PL}. We can clearly see that $P_L(B^0\to K_0^*(1430)e^+e^-)$ and $P_L(B^0\to K_0^*(1430)\mu^+\mu^-)$ are close to $-1$ expect those close to the end points of $q^2_{\rm min}$ and $q^2_{\rm max}$. And $P_L(B^0\to K_0^*(1430)\tau^+\tau^-)$ is in the range of $-0.5$ to $0$, and our results are in good agreement with LCSR 2008~\cite{Wang:2008da} in the Fig.~\ref{fig:PL}. At the same time, we can obtain the average numerical results of integrated longitudinal lepton polarization asymmetry by the following formula
\begin{align}
\langle A_{P_L} \rangle = \int^{s_{\rm max}}_{s_{\rm min}} A_{P_L} (s) ds,
\label{eq:AL}
\end{align}
which are shown in Table~\ref{table:AL}, respectively. Our predictions are in a good agreement with Refs.~\cite{Wang:2008da,Chen:2007na}. Despite the differences in the results of TFFs, the results of integrated longitudinal
lepton polarization asymmetry are the same, $i.e.,$ the TFFs have little effect on this asymmetry. Meanwhile, because the efficiency of tau lepton detectability is too low, we are currently unable to measure the $\tau$ lepton polarization.

\begin{table}[t]
\begin{center}
\renewcommand{\arraystretch}{1.5}
\footnotesize
\caption{The average numerical results of integrated longitudinal lepton polarization asymmetries $\langle A_{P_L} \rangle$ for $B^0\to K_0^*(1430)\ell^+\ell^-$ with $\ell=e,\mu,\tau$, and the predictions of LCSR 2008~\cite{Wang:2008da} and LFQM 2007~\cite{Chen:2007na} are used to compare with our results.}.
\label{table:AL}
\begin{tabular}{l llll}
\hline
&$B^0\to K_0^* e^+e^-$&$B^0\to K_0^* \mu^+\mu^-$&$B^0\to K_0^* \tau^+\tau^-$\\
\hline
This work & $-0.99$   &$-0.96$   &$-0.03$  \\
LCSR 2008~\cite{Wang:2008da}  & $-0.99\pm0.0$   &$-0.96\pm0.0$  &$-0.03^{+0.00}_{-0.01}$\\
LFQM 2007~\cite{Chen:2007na} & $-0.97$   &$-0.95$  &$-0.03$  \\
\hline
\end{tabular}
\end{center}
\end{table}

\section{Summary}\label{Sec:IV}
In this paper, in order to obtain precise numerical analysis results of observations to test SM, we have investigated the rare decay $B\to K_0^{\ast}(1430)\ell^+\ell^- (\nu\bar{\nu})$ induced by the FCNC transition of $b\to s\ell^+\ell^- (\nu \bar{\nu})$ with $\ell=(e,\mu,\tau)$, which are quite sensitive to SM. Firstly, based on the scenario 2 that the $K_0^*(1430)$ is viewed as the ground state of $s\bar{q}$ and $q\bar{s}$, we calculate the $B^0
\to K_0^*(1430)$ TFFs $f^{B^0 \to K_0^*}_{\pm,{\rm T}}(q^2)$ with LCSR method up to NLO accurancy. To make the calculation more accurate, we adopte the improved $\phi_{2;K_0^*}(x,\mu)$, $ie.$ Eq.~\eqref{eq:leading-twist} constructed by a new LCHO model and twist-3 DAs $\phi_{3;K_0^*}^p(x,\mu)$, $\phi_{3;K_0^*}^\sigma(x,\mu)$ determined by QCD background field theory from our previous work~\cite{Huang:2022xny,Han:2013zg} and use the SSE to extrapolate TFFs for $f^{B^0 \to K_0^*}_{\pm,{\rm T}}(q^2)$ to the whole kinematical region $0< q^2<(m_{B^0}-m_{K_0^*})^2$. The numerical results for TFFs at large recoil region are listed in Table~\ref{table:TFFsvalue} and our predictions of TFFs are consistent with other LCSR predictions within errors. The behaviors of TFFs is showed in Fig.~\ref{fig:TFFs}, in which we disscussed the results of TFFs in the applicable region of LCSR method, $i.e.$, low and intermediate $q^2$-region, $q^2 \in[0,8]~\rm{GeV^2}$. And our predictions are in a agreement with LCSR 2013~\cite{Han:2013zg} and LCSR 2015~\cite{Wang:2014vra} in this region.

Then, we utilize TFFs to calculated the differential decay widths, branching fractions and longitudinal lepton polarization asymmetries. The Fig.~\ref{fig:differenal DW} shows the differential decay widths and our results of the electron and muon pair channel have different agreement with other theory group in various $q^2$-region due to the difference of TFFs.  But the tauon pair channel have a well agreement with LCSR 2008~\cite{Wang:2008da}, LCSR 2015~\cite{Wang:2014upa} and pQCD~\cite{Li:2008tk} in whole $q^2$-region within errors. And the $B^0\to K_0^*(1430) \nu\bar{\nu}$ differential decay width is in a agreement with LCSR 2015~\cite{Wang:2014upa} in whole $q^2$-region. Meanwhile we give the results of total decay width of $B^0\to K_0^*(1430) \ell^+\ell^-$ and $B^0\to K_0^*(1430) \nu\bar{\nu}$, which help us to further calculate the branching fraction. The predictions of $B^0\to K_0^*(1430) \ell^+\ell^-$ and $B^0\to K_0^*(1430) \nu\bar{\nu}$ are of the order $10^{-7}\sim10^{-8}$ and $10^{-6}$ , respevtively, which have a great possibility that can be observed in future experiments. Our results are consistent with most of theoretical predictions. The rationality of these results also indirectly proves the feasibility of the LCHO model $\phi_{2;K_0^*}(x,\mu)$ in these decay process.

Finally, the longitudinal lepton polarization asymmetries of $B^0\to K_0^*(1430) \ell^+\ell^-$ are shown in Fig.~\ref{fig:PL} and the numerical results of integrated longitudinal lepton polarization asymmetries are listed in Table~\ref{table:AL}. The influence of TFFs on this observation are not significant and our predictions are in good agreement with the results of other LCSR~\cite{Wang:2008da} and LFQM~\cite{Chen:2007na}. Compared with tauon, the electron and muon lepton polarization asymmetries have greater hope to be detected. These theoretical predictions in these rare decay processes are very helpful for us to test the SM and understand the strong interaction. We believe that $B$-meson decays into the $P$-wave scalar meson $K_0^*(1430)$ will soon be measured in the experiment in the future.
\\
\section{Acknowledgments}

Hai-Bing Fu would like to thank the Institute of High Energy Physics of Chinese Academy of Sciences for their warm and kind hospitality. This work was supported in part by the National Natural Science Foundation of China under Grant No.12265009 and 12265010, the Project of Guizhou Provincial Department of Science and Technology under Grant No.ZK[2023]024.

\appendix

\section{Specific expression of NLO for perturbative $\mathcal{O}(\alpha_s)$ correction to the twist-2 terms}\label{Appendix}

The specific expression for the perturbative $\mathcal{O}(\alpha_s)$ corrections to the twist-2 terms from Ref~\cite{Wang:2014vra} are
\begin{widetext}
\begin{eqnarray}
\rho_+^{\alpha_s} &=&\frac{C_F\alpha_s}{2\pi} \int^1_{\bar{\Delta}} du \phi(u) \Bigg\{\frac{1}{r_2} + \frac{2\bar{u}}{1-r_2} - \frac{\log r_2 - 2\log(r_2-1)-1}{r_2-r_1}\nonumber\\
&+& \Theta(\rho -1) \Bigg[\frac{1-\rho}{2\rho^2}- \frac{4\log(\rho-1)}{1-\rho}\Big|_+ + \frac{2\log \rho}{1-\rho}\Big|_+ - \frac{2}{\rho} - \frac{1-r_2}{r_2}\frac{1}{1-\rho}\Big|_+ +\frac{\log \rho +1 -2\log(\rho-1)}{u(r_2-r_1)}\Bigg]\nonumber\\
&+& \Theta(1-\rho)\Bigg[ -\frac{1-r_2}{r_2}\frac{1}{1-\rho}\Big|_+\Bigg] \nonumber\\
&+& \delta(1-\rho)\Bigg[-\frac{5}{2}\log \frac{m_b^2}{\mu^2} + 2\log^2(r_2-1)+2{\rm Li}_2(1-r_2)-\frac{\pi^2}{3}+6  \nonumber\\
&+& \frac{1-r_2}{r_2}[2\log (r_2-1)-\log(1-r_1)]+\Big(1-\log\frac{m_b^2}{\mu^2}\Big)\Big(1+\frac{1}{r_2-r_1}\frac{\mathop{d}\limits^{\leftarrow}}{du}\Big) \nonumber\\
&+&  \frac{2F(r_1)-(1-r_1)G(r_1)-2}{u(r_2-r_1)}\Bigg]\Bigg\}\nonumber\\
\rho_{+-}^{\alpha_s}&= &\frac{{\rm Im}\tilde{\Pi}_{+-}(q^2,s)}{\pi}\nonumber\\
&=&\frac{C_F\alpha_s}{2\pi} \int^1_{\bar{\Delta}} du \phi(u) \Bigg\{ \frac{1}{r_2}-\Theta(\rho-1)\frac{\rho-1}{2u\rho^2}+\delta(1-\rho)\frac{(1-r_1)F(r_1)}{2ur_1^2}\Bigg\}\nonumber\\
\rho_{T}^{\alpha_s}&=&\frac{{\rm Im}\tilde{\Pi}_{T}(q^2,s)}{\pi}\nonumber\\
&=&\frac{C_F\alpha_s}{2\pi m_b} \int^1_{\bar{\Delta}} du \phi(u) \Bigg\{ \frac{2\log(r_2-1)-\log r_2+1}{r_2-r_1}-\frac{r_2-1}{r_2(r_2-r_1)} +\frac{\bar{u}}{r_2}+\frac{2\bar{u}}{1-r_2}\nonumber\\
&+&\Theta(\rho-1)\Bigg[-\frac{3}{2\rho}-\frac{4\log(\rho-1)}{1-\rho}\Big|_+ +\frac{2\log \rho}{1-\rho}\Big|_+ +\frac{\log \rho -2\log(\rho-1)-1}{u(r_2-r_1)}\nonumber\\
&+&\Big(3-\frac{1}{\rho}\Big)\frac{1}{u(r_2-r_1)}+\frac{5\rho-1}{2\rho^2}\frac{1}{1-\rho}\Big|_+ \Big] +\delta(1-\rho)\Bigg[ \frac{9}{2}-\frac{\pi^2}{3} + 2\log^2(r_2-1)\nonumber\\
&+& 2{\rm Li}_2(1-r_2)-\frac{3}{2}\log \frac{m_b^2}{\mu^2}+\Big(3F(r_1)-(1-r_1)G(r_1)-2-\frac{F(r_1)}{r_1}\Big)\frac{1}{u(r_2-r_1)} \nonumber\\
&-&2\log(r_2-1)-3\log \frac{m_b^2}{\mu^2}\frac{1}{r_2-r_1}\frac{\mathop{d}\limits^{\leftarrow}}{du}
+\frac{9}{2}\frac{1}{r_2-r_1}\frac{\mathop{d}\limits^{\leftarrow}}{du}-\frac{1}{2\rho}\frac{1}{r_2-r_1}\frac{\mathop{d}\limits^{\leftarrow}}{du}\Bigg]\Bigg\}
\end{eqnarray}
\end{widetext}
where
\begin{align}
\bar{\Delta}&=\frac{m_b^2}{s_0-q^2},\nonumber\\
r_1=\frac{q^2}{m_b^2},~r_2&=\frac{(p+q)^2}{m_b^2},~\rho=r_1+u(r_2-r_1),\nonumber\\
\phi(\rho)\frac{f(\rho)}{1-\rho}\Big|_+ &= \frac{f(\rho)}{1-\rho}\Big[\phi(\rho)-\phi(1)\Big],\nonumber\\
F(\rho)&=(1-\rho)\log(1-\rho)+\rho,\nonumber\\
G(\rho)&=\log^2(1-\rho)+Li_2(\rho)+\log(1-\rho).
\end{align}

\end{document}